\documentclass[twocolumn,a4,showpacs,preprintnumbers,amsmath,amssymb]{style}
\usepackage{graphicx}
\usepackage{epsfig}
\usepackage{amsmath,amssymb}
\usepackage{bm}
\usepackage{a4wide}
\usepackage{amsmath}
\usepackage{amssymb}
\usepackage{ifthen}
\usepackage{epsfig}
\usepackage{epsfig}
\usepackage{longtable}

\usepackage{graphicx}
\usepackage{dcolumn}
\usepackage{bm}
\usepackage{mathptmx, courier, pifont}
\usepackage[scaled=1.0]{helvet}
\usepackage[T1]{fontenc}
\usepackage{textcomp}
\usepackage{color}
\usepackage{colordvi}

\usepackage{rotating}
\usepackage{rotating}

\def\bra{\,<\!} \def\ket{\!>\,} \def\ack{\,|\,}
\usepackage[ps2pdf,colorlinks=true]{hyperref}

\begin{document}

\title{Microscopic nuclear structure models and methods :
Chiral symmetry, Wobbling motion and $\gamma-$bands}

\author{ Javid A. Sheikh\email{sjaphysics@gmail.com}\,  , Gowhar H. Bhat\email{gwhr.bhat@gmail.com} ,  Waheed A. Dar  \\
\it Department of Physics, University of Kashmir, Srinagar, 190 006, India  \\
Sheikh Jehangir, Prince A. Ganai\\
\it Department of Physics, National Institute of Technology, Srinagar, 190 006, India \\
}
\date{}

\maketitle

\begin{abstract}
A systematic investigation of the nuclear observables related to the
triaxial degree of freedom is presented using the 
multi-quasiparticle triaxial projected shell model
(TPSM) approach. These properties correspond to the observation of $\gamma$-bands,
chiral doublet bands and the wobbling mode. In the TPSM
approach, $\gamma$-bands are built on each quasiparticle
configuration and it is demonstrated that some observations in
high-spin spectroscopy that have remained unresolved for quite some time
could be explained by considering $\gamma$-bands based on two-quasiparticle
configurations.  It is shown in some Ce-, Nd- and Ge-isotopes
that the two observed aligned or s-bands originate from the same intrinsic
configuration with one of them as the $\gamma$-band based on
a two-quasiparticle configuration. In the present work, we have also
performed a detailed study of $\gamma$-bands observed up to the highest 
spin in Dysposium, Hafnium, Mercury and Uranium
isotopes.  Furthermore, several measurements
related to chiral symmetry breaking and wobbling motion have been
reported recently. These phenomena, which are possible only for triaxial nuclei,
have been investigated using the TPSM approach. It is shown that
doublet bands observed in lighter odd-odd Cs-isotopes can be 
considered as candidates for chiral symmetry breaking.
Transverse wobbling motion recently observed in $^{135}$Pr has also been investigated
and it is shown that TPSM approach provides a reasonable description
of the measured properties. 
\end{abstract}


\section{Introduction}
\label{Sect.01}
Atomic nucleus is one of the most fascinating quantum many-body systems
that depicts a rich variety of shapes and structures. At the same time,
it is also one of the most challenging problems in physics to
investigate theoretically. The number of particles is not too large as
in condensed matter physics so that  statistical tools become 
applicable and also it is not too small such that few-body techniques 
can be employed for nuclei across the periodic table. The phenomenal
progress in understanding the properties of nuclei has been achieved
using phenomenological models and methods. These phenomenological
models are primarily based on empirical observations and have played a pivotal
role in unraveling the intrinsic structures of atomic nuclei
 \cite{Rc20,Ku84,Mg49,GS74,VS76}.  

The pioneering work of Bohr, Mottelson and Rainwater laid the
foundation of the phenomenological models in nuclear physics. It was
demonstrated that properties of atomic nuclei can be elucidated by
considering rotational, vibrational and single-particle motion as three basic degrees
of freedom and led to the development of the collective model in
sixties and seventies \cite{BM75,SN55}. 
This model is even being used today with the parameters 
determined through microscopic approaches rather than following the
empirical route.

The nuclear physics research is going through a renaissance with the
state-of-the-art tools and techniques being developed to probe the
wealth of nuclear properties.  The availability of the leadership
computing facilities has made it possible to apply {\it {Ab-initio}} methods
to lighter mass region with remarkable success. 
On the other hand, the
density functional approach is now widely used to explore the
ground-state properties all across the nuclear landscape \cite{je12,WK65,WK64}. 
In recent
years, the progress achieved in applying these modern tools is quite remarkable and it is
expected that it would be possible to apply these techniques to
investigate a broad spectrum of nuclear properties all across the nuclear periodic
table \cite{JD94,MB03} in a forseeable future. However, at the moment, these models have limitations and
cannot be employed to investigate, for instance, the rich-band 
structures observed in deformed nuclei. 

In the absence of a fully microscopic theory, semi-microscopic models have been developed to study the properties
of band structures in medium and heavy mass nuclei. In this class of
models, the triaxial projected shell model (TPSM)  approach has been
demonstrated to correlate the high-spin data of well deformed and
transitional data with remarkable success \cite{SH99}. 
The purpose of the present
work is to provide an overview of the recent applications of the
TPSM approach to a wide range of nuclear properties  \cite{GH08,JG09,SB10,JG11,Yeo11,GJ12,JG12,bh13,bh13,GH14,bh14,bh14a,bh14b,bh15,bh15a,bh15b}. 
We also report new results on the observation of $\gamma$
bands based on excited configurations. 
Furthermore, we shall present a systematic investigation of $\gamma-$bands
observed up to the highest spin in Dysposium, Hafnium, Mercury and Uranium
isotopes. 
The manuscript is organised
in the following manner. In the next section, a few details of the TPSM
approach are provided and some technical aspects of the model are 
included in the appendix. In section 3, the results of the calculations
for $\gamma$-, chiral- and wobbling- bands are displayed and discussed.
Finally, the work presented in this manuscript is summarized and concluded in section 4.


\section{
Outline of the Triaxial Projected Shell Model Approach}
\label{Sect.02}

The basic philosophy of the TPSM approach is similar to the spherical
shell model model (SSM) with the only difference that deformed basis
are employed for  diagonalizing  the shell model Hamiltonian
rather than the spherical one. The deformed basis are constructed
by solving the triaxial Nilsson potential with optimum 
quadrupole deformation parameters of $\epsilon$ and $\epsilon'$.
In principle, the deformed basis can be constructed with arbitrary
deformation parameters, however, the basis are constructed with 
expected or known deformation parameters (so called optimum) for a given system
under consideration. These deformation values lead to an accurate
Fermi surface and it is possible to 
choose a minimal subset of the basis states around the Fermi
surface for a realistic description of a given system. The Nilsson basis
states are then transformed to the quasiparticle space using the simple
Bardeen-Cooper-Schriefer (BCS) ansatz for treating the pairing
interaction.

As the deformed basis are defined in the intrinsic frame of reference
and don't have well defined angular-momentum, in the second stage these basis are
projected onto states with well defined angular-momentum using the
angular-momentum projection technique \cite{ring80,HS79,HS80}. 
The three dimensional angular-momentum projection
operator is given by 
\begin{equation}
\hat P ^{I}_{MK}= \frac{2I+1}{8\pi^2}\int d\Omega\, D^{I}_{MK}
(\Omega)\,\hat R(\Omega),
\label{Anproj}
\end{equation}
with the rotation operator 
\begin{equation}
\hat R(\Omega)= e^{-i\alpha \hat J_z}e^{-i\beta \hat J_y}
e^{-i\gamma \hat J_z}.\label{rotop}
\end{equation}
 Here, $''\Omega''$ represents a set of Euler angles 
($\alpha, \gamma = [0,2\pi],\, \beta= [0, \pi]$) and the 
$\hat{J}'s$ are angular-momentum operators.
The projected basis states
considered in the present work for the even-even system are composed
of vacuum, two-proton, two-neutron and two-proton plus two-neutron
configurations, i.e.,
\begin{eqnarray}
\{ \hat P^I_{MK}\left|\Phi\right>, ~\hat P^I_{MK}~a^\dagger_{p_1}
a^\dagger_{p_2} \left|\Phi\right>, ~\hat P^I_{MK}~a^\dagger_{n_1}
a^\dagger_{n_2} \left|\Phi\right>,  \nonumber \\~\hat
P^I_{MK}~a^\dagger_{p_1} a^\dagger_{p_2} a^\dagger_{n_1}
a^\dagger_{n_2} \left|\Phi\right> \}, \label{basis}
\end{eqnarray}
where $\left|\Phi\right>$ in (\ref{basis}) represents the triaxial qp
vacuum state. The above basis space used for the even-even system
is sufficient to study nuclei up to second band crossing region and in
the rare-earth region this means approximately up to spin,
I=24~$\hbar$. For odd-proton (neutron) systems, the basis space is composed of 
one-quasiproton (quasineutron) and two-quasineutrons (quasiprotons).
In the case of odd-odd nuclei, the basis space is simply
one-quasiproton coupled to one-
quasineutron. 
 
The advantage of the TPSM approach as compared to the other
approaches, for instance the cranking approach, is that not only 
the yrast band but also the rich excited
band structures can  be investigated. As a matter of fact, the
major focus of the present work is to study the $\gamma$-bands
which form the first excited band in many transitional nuclei. 
The  Nilsson triaxial quasiparticle states don't have well defined 
projection along the symmetry axis, $\Omega$ and are a superposition
of these states. For instance, the triaxial self-conjugate vacuum state is a 
superposition of K = 0, 2, 4,...states - only even-states 
are possible  due to symmetry requirement \cite{YK00}. 
For the symmetry operator, 
$\hat S = e^{-\imath \pi \hat J_z}$, we have 
\begin{equation}
\hat P^I_{MK}\ack\Phi\ket = \hat P^I_{MK} \hat S^{\dagger} \hat 
\ack\Phi\ket = e^{\imath \pi (K-\kappa)} \hat P^I_{MK}\ack\Phi\ket,
\end{equation}
 where, $\hat S\ack\Phi\ket = e^{-\imath \pi \kappa}\ack\Phi\ket$, and
$\kappa$ characterizes the intrinsic states.
For the self-conjugate vacuum state $\kappa=0$ and,
therefore, it follows from the above equation that only $K=$ even,
values are permitted for this state. For 2-qp states, the possible
values for $K$-quantum number are both even and odd depending on
the structure of the qp state. For the 2-qp state formed from the
combination of the normal and the time-reversed states, $\kappa = 0$
and again only $K=$ even values are permitted. For the
combination of the two normal states, $\kappa=1$, and only $K=$ odd
states are allowed.

The projected
states for a given configuration that constitute a rotational band
are obtained by specifying the corresponding  K-value in the angular-momentum
projection operator. The projected states from K = 0, 2 and 4 correspond
to ground-, $\gamma-$ and $\gamma\gamma-$bands, respectively.
As stated earlier, for two-quasiparticle states, both even- and odd-K values are
permitted depending on the signature of the two quasiparticle states.
In this description, the aligning states that cross the ground-state
band and lead to upbend or backbend phenomenon have K = 1. The
projection from the same quasiparticle intrinsic state with K = 3 is the 
$\gamma$-band built on these two quasiparticle state. The $\gamma$-bands
built on the two-quasiparticle states shall form one of the major
focal issues of the present work and shall be discussed in detail in the
next section.

In the third and the final stage of the TPSM analysis, the projected basis are
employed to diagonalize the shell model Hamiltonian. The model
Hamiltonian consists of pairing and quadrupole-quadrupole interaction
terms, i.e.,
\begin{eqnarray}
\hat H =  \hat H_0 &-&   {1 \over 2} \chi \sum_\mu \hat Q^\dagger_\mu
\hat Q^{}_\mu - G_M \hat P^\dagger \hat P \nonumber \\ & -& G_Q \sum_\mu \hat
P^\dagger_\mu\hat P^{}_\mu . \label{hamham}
\end{eqnarray}
In the above equation, $\hat H_0$ is the spherical single-particle
Nilsson Hamiltonian \cite{Ni69}. The parameters of the Nilsson
potential are fitted to a broad
range of nuclear properties and is quite appropriate to employ it as a
mean-field potential.
The QQ-force strength, $\chi$, in Eq. (\ref{hamham}) is related to
the quadrupole deformation $\epsilon$ as a result of the
self-consistent HFB condition and the relation is given by
\cite{KY95}:
\begin{equation}
\chi_{\tau\tau'} =
{{{2\over3}\epsilon\hbar\omega_\tau\hbar\omega_{\tau'}}\over
{\hbar\omega_n\left<\hat Q_0\right>_n+\hbar\omega_p\left<\hat
Q_0\right>_p}},\label{chi}
\end{equation}
where $\omega_\tau = \omega_0 a_\tau$, with $\hbar\omega_0=41.4678
A^{-{1\over 3}}$ MeV, and the isospin-dependence factor $a_\tau$ is
defined as
\begin{equation}
a_\tau = \left[ 1 \pm {{N-Z}\over A}\right]^{1\over 3},\nonumber
\end{equation}
with $+$ $(-)$ for $\tau =$ neutron (proton). The harmonic
oscillation parameter is given by $b^2_\tau=b^2_0/a_\tau$ with
$b^2_0=\hbar/{(m\omega_0)}=A^{1\over 3}$ fm$^2$. 
 The monopole pairing strength $G_M$ (in MeV)
is of the standard form
\begin{eqnarray}
&&G_M = {{G_1 - G_2{{N-Z}\over A}}\over A} ~{\rm for~neutrons,}~~~~\\\nonumber
&&G_M = {G_1 \over A} ~{\rm for~protons.} \label{pairing}
\end{eqnarray}
In the present calculation, we choose  $G_1$ and $G_2$
such that the observed odd-even mass difference is reproduced
in the mass region. The values $G_1$ and $G_2$ vary depending
on the mass region and shall be mentioned in the discussion
of the results. The above choice of $G_M$ is appropriate for the
single-particle space employed in the model, where three major
shells are used for each type of nucleon. The quadrupole pairing strength $G_Q$ is
considered to be proportional to $G_M$ and the proportionality
constant being fixed as 0.18. These interaction strengths are
consistent with those used in our earlier studies
\cite{GH08,JG09,SB10,JG11,GJ12,YK00,KY95,JY01,js01,rp01,Ch12,YS08}.

It is shown in the appendix that the projection formalism outlined above
can be transformed into a diagonalization problem following the
Hill-Wheeler prescription. The Hamiltonian in Eq. (\ref{hamham}) is diagonalized using the projected 
basis of Eq. (\ref{basis}). The obtained wavefunction can be written as
\begin{equation}
\psi^{\sigma}_{IM} = \sum_{K,\kappa}a^{\sigma}_{\kappa}\hat P^{I}_{MK}| 
~ \Phi_{\kappa} \ket.
\label{17}
\end{equation}
Here, the index $\sigma$ labels the states with same angular momentum
and $\kappa$ the basis states. In Eq. (\ref{17}),
$a^{\sigma}_{\kappa}$ are the amplitudes of the basis states
$\kappa$. These wavefunction are used to 
calculate the electromagnetic transition probabilities.   The reduced electric quadrupole transition probability $B(E2)$ from an initial state 
$( \sigma_i , I_i) $ to a final state $(\sigma_f, I_f)$ is given by \cite {su94}
\begin{equation}
B(E2,I_i \rightarrow I_f) = {\frac {e^2} {2 I_i + 1}} 
| \bra \sigma_f , I_f || \hat Q_2 || \sigma_i , I_i\ket |^2 .
\label{BE22}
\end{equation}
As in our earlier publications \cite{JG12,bh13,GH14,bh14,bh14a}, we have used the effective charges
of 1.6e for protons and 0.6e for neutrons. 
The reduced magnetic dipole transition probability
$B(M1)$ is computed through
\begin{equation}
B(M1,I_i \rightarrow I_f) = {\frac {\mu_N^2} {2I_i + 1}} | \bra \sigma_f , I_f || \hat{\mathcal M}_1 ||
\sigma_i , I_i \ket | ^2 , 
\label{BM11}
\end{equation}
where the magnetic dipole operator is defined as  
\begin{equation}
\hat {\mathcal {M}}_{1}^\tau = g_l^\tau \hat j^\tau + (g_s^\tau - g_l^\tau) \hat s^\tau . 
\end{equation}  
Here, $\tau$ is either $\nu$ or $\pi$, and $g_l$ and $g_s$ are the orbital and the spin gyromagnetic factors, 
respectively. 
In the calculations
we use for $g_l$ the free values and for $g_s$ the free values 
damped by a 0.85 factor, i.e.,
\begin{eqnarray}
&&g_l^\pi = 1, ~~~ 
g_l^\nu = 0, ~~~   \nonumber\\
&&g_s^\pi =  5.586 \times 0.85,\nonumber\\
&& g_s^\nu = -3.826 \times 0.85.
\end{eqnarray}
The reduced matrix element of an operator $\hat {\mathcal {O}}$ ($\hat {\mathcal {O}}$ is either
$\hat {Q}$ or $\hat {\mathcal {M}}$) can be expressed as
\begin{eqnarray}
&&\bra \sigma_f , I_f || \hat {\mathcal {O}}_L || \sigma_i , I_i\ket 
\nonumber \\ & =& 
\sum_{\kappa_i , \kappa_f} {f_{I_i \kappa_i}^{\sigma_i}} {f_{I_f \kappa_f}^{\sigma_f}}
 \sum_{M_i , M_f , M} (-)^{I_f - M_f}  \nonumber \\&\times&
\left( \begin{array}{ccc}
 I_f & L & I_i \\
-M_f & M & M_i 
\end{array} \right) \nonumber \\
 &\times&  \bra \phi_{\kappa_f} | {\hat{P}^{I_f}}_{K_{\kappa_f} M_f} \hat {\mathcal {O}}_{LM}
\hat{P}^{I_i}_{K_{\kappa_i} M_i} | \phi_{\kappa_i} \ket  \nonumber \\
&=& 2 \sum_{\kappa_i , \kappa_f} {a_{I_i \kappa_i}^{\sigma_i}} {a_{I_f \kappa_f}^{\sigma_f}}\sum_{M' , M''} (-)^{I_f - K_{\kappa_f}} {(2I_f + 1)}^{-1}\nonumber\\
&\times&   \left( \begin{array}{ccc}
 I_f & L & I_i \\
-M_{\kappa_f} & M' & M'' 
\end{array} \right)~\int d\Omega \,D_{M''\kappa_{\kappa_i}}(\Omega)\nonumber\\
&\times&\bra \Phi_{\kappa_f}|\hat {\mathcal {O}}_{LM'}\hat R(\Omega)|\Phi_{\kappa_i}\ket.
\label{beeee2}
\end{eqnarray}
In the above expression, the symbol $(~~~~~~)$ denotes a 3j-coefficient.
\begin{table}[htb]
\begin{center}
\caption{Axial and non-axial quadrupole deformation values $\epsilon$
and $\epsilon'$ employed in the TPSM calculation for  Ce and Nd isotopes. Axial 
deformations $\epsilon$ have been considered from \cite{MN95} and
nonaxial values are chosen in such a way that band heads of the $\gamma-$
bands are reproduced.}
\vspace{0.9cm}
\begin{tabular}{cccc}
\hline
 A & $\epsilon$ & $\epsilon'$ & $\gamma$\\ \hline
  $^{132}$Ce & 0.183 & 0.100 & 29 \\ 
  $^{134}$Ce & 0.150 & 0.100 & 34 \\ 
  $^{134}$Nd & 0.200 & 0.120 & 31\\ 
  $^{136}$Nd & 0.158 & 0.110 & 35\\ 
  $^{138}$Nd & 0.170 & 0.110 & 33\\ \hline
\end{tabular}\label{defcd}
\end{center}
\end{table}
\begin{figure}[thb]
\begin{center}
  \includegraphics[width=0.9\columnwidth]{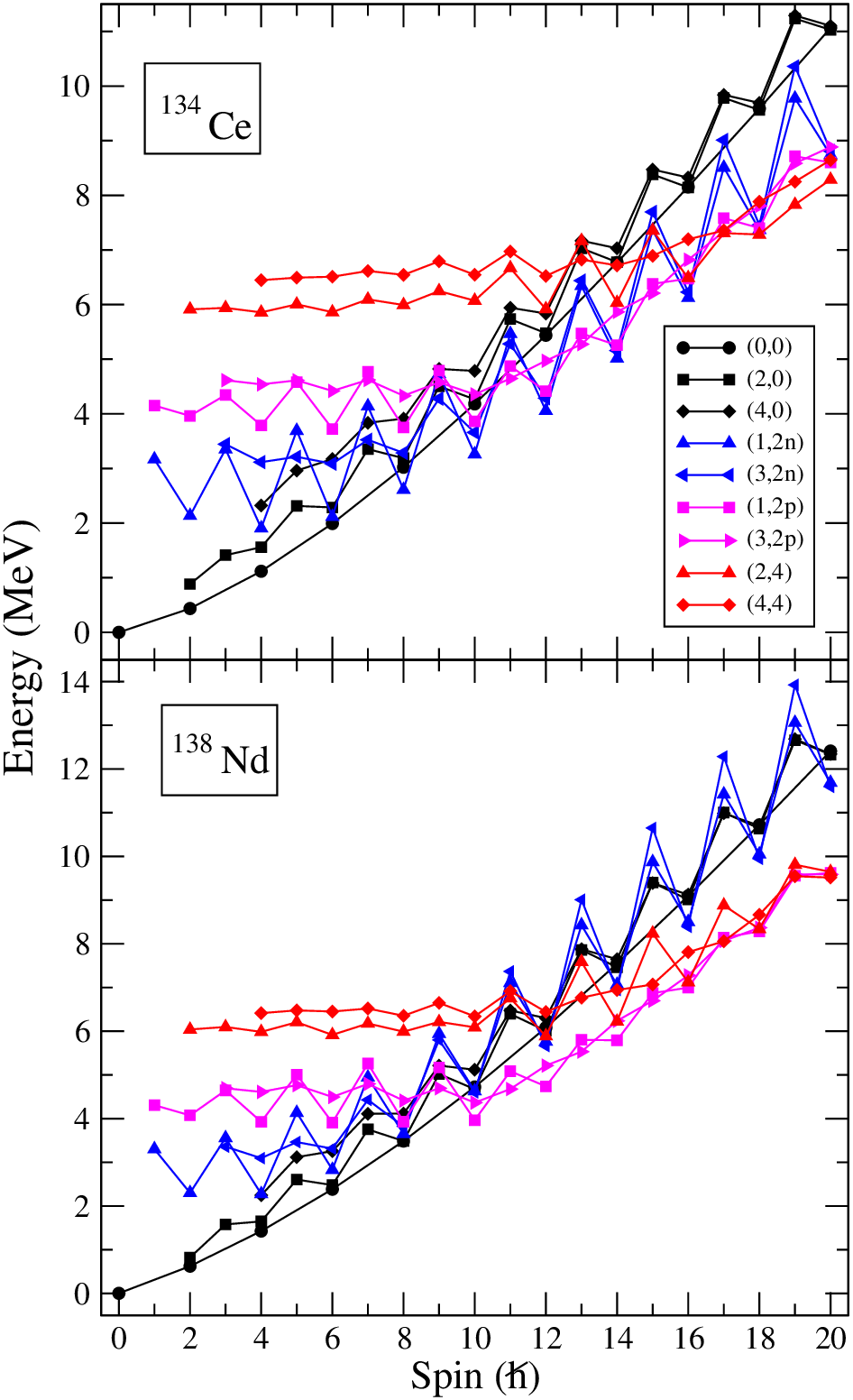}
\end{center}
\caption{\label{fig.ce1}  (Color online) Band diagrams for $^{134}$Ce and $^{138}$Nd isotopes 
as representative examples. The labels
($K$,n-qp) indicate the $K$-value and the quasiparticle character of
the configuration, for instance, $(3,2n)$ corresponds to the
2n-aligned $\gamma$-band built on 2n-aligned state.
}
\end{figure}

\begin{figure}[thb]
\begin{center}
  \includegraphics[width=0.9\columnwidth]{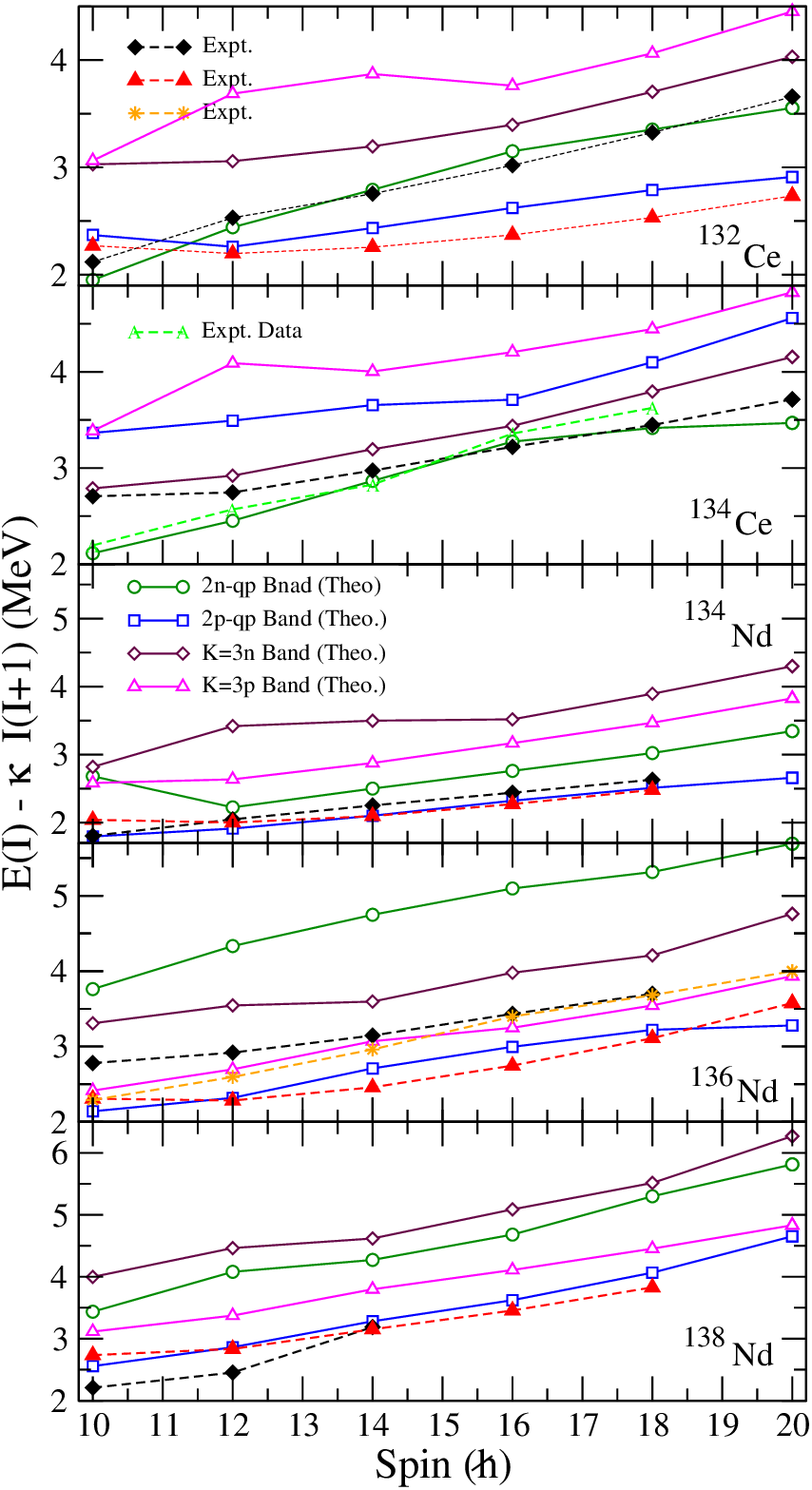}
\end{center}
\caption{\label{fig.ce2}  (Color online) Theoretical bands with the dominant component from (1, 2n), (1, 2p), (3, 2n), and (3, 2p).
Only spin range from I = 10 to 20 is shown for which these bands are low in energy. Available data in
$^{132,134}$Ce and $^{134,136,138}$Nd are compared with the calculated
results. Data has been taken from Refs. 
\cite{132cs,134cs,134nd,136nd,138nd}.
}
\end{figure}
\section{Results and Discussions}
In the past few years, TPSM approach has been used quite extensively
to shed light on some of the
outstanding issues related to the triaxility in atomic nuclei and
in the present section we shall provide a brief overview of these
investigations and 
some new results obtained recently shall also be
presented and discussed. This section is divided into 
subsections, discussing various aspects and implications on the presence of
triaxial deformations in atomic nuclei. 
\subsection{Observation of the $\gamma$-bands based on
  two-quasiparticle states}

$\gamma$-bands are observed in most of the transitional nuclei all
across the nuclear periodic table and have been studied using various
theoretical approaches and methods
\cite{Pj91,Mm87,GY81,LP88,JP88,MK87,Er92,BJ93,BS94,AI75,YA86,CL87,Ga96,EM70, JH80,AS58}.
 In phenomenological models, these bands are
interpreted as emerging due to vibrational motion in the $\gamma$ degree of
freedom of the nuclear deformation
\cite{Gu95,WA93,WP93,BJ91,MK88,VG81,AB82,DG94,
  Ns99,Rf97,As58,Cf96,Ti82}.
\begin{figure}[htb]
 \centerline{\includegraphics[trim=0cm 0cm 0cm
0cm,width=0.45\textwidth,clip]{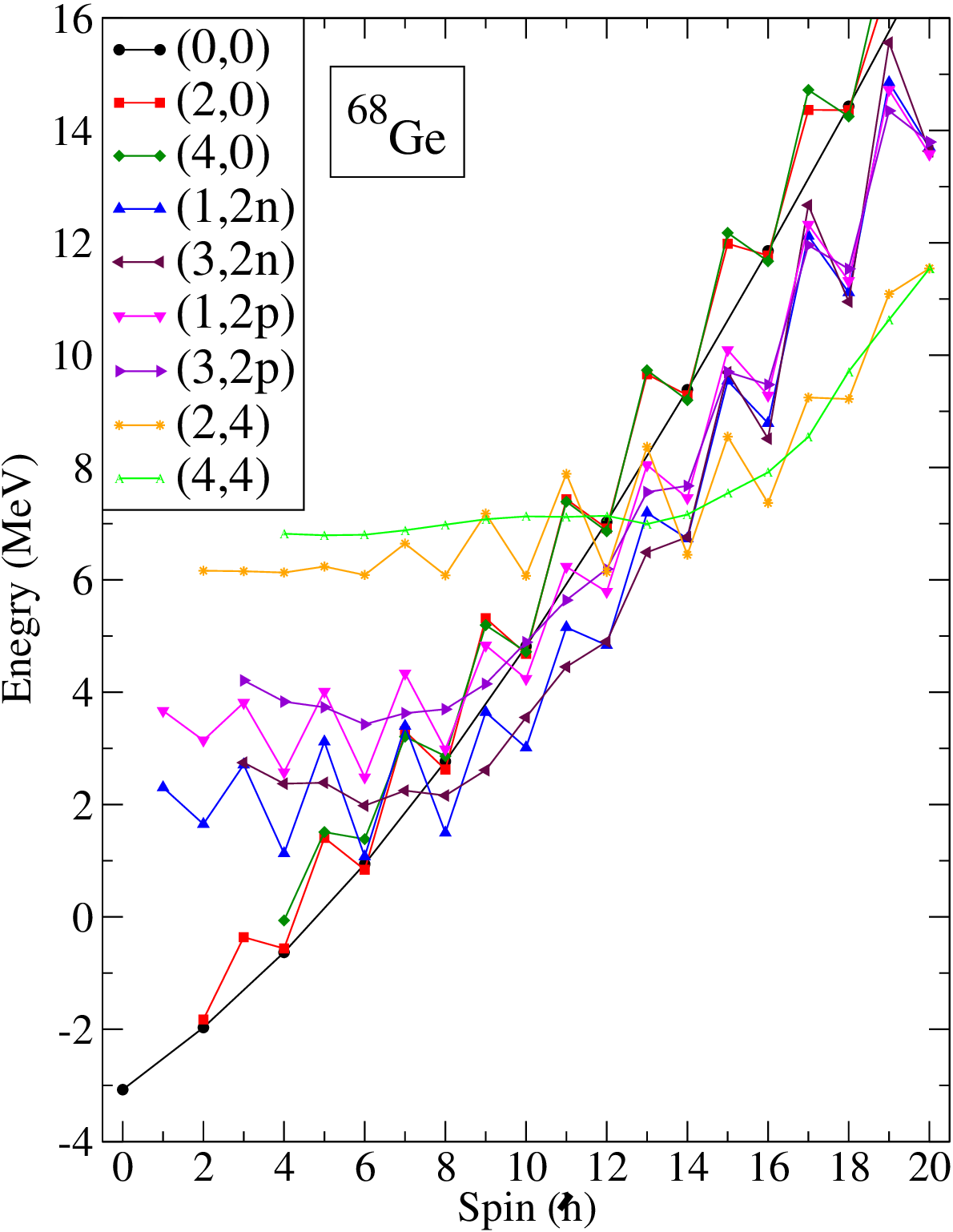}}
\caption{(Color online) Band diagram for $^{68}$Ge isotope. The labels
($K$,n-qp) indicate the $K$-value and the quasiparticle character of
the configuration, for instance, $(3,2n)$ corresponds to the
2n-aligned $\gamma$-band built on  2n-aligned state K=1 quasiparticle configuration.
}\label{figtpsm1}
\end{figure}
%
\begin{figure}[thb]
\begin{center}
  \includegraphics[width=0.9\columnwidth]{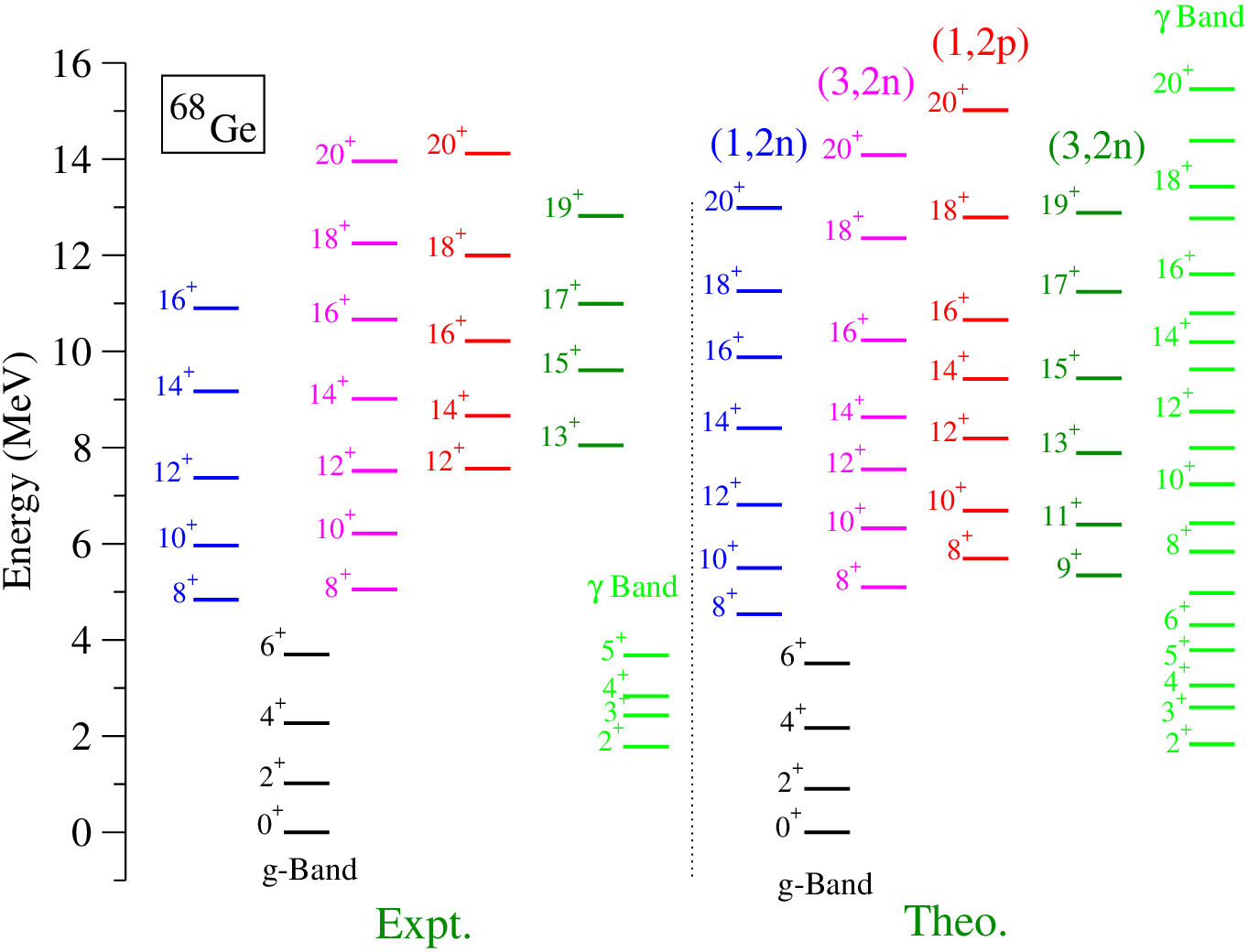}
\end{center}
\caption{ (Color online) Comparison of the measured energy levels of  $^{68}$Ge nucleus  and the  results of TPSM calculations. Data has been taken from Refs. \cite{68Ge}.
}\label{figtpsm2}
\end{figure}
In the microscopic TPSM
approach, $\gamma$-band structure results from the projection of the
K = 2 component of the triaxial vacuum state. $\gamma$-bands are also
possible based on the multi-quasiparticle triaxial states apart from the
vacuum configuration and have not been studied as most of the
models consider vacuum configuration only. The basis of TPSM approach has been
enlarged to include multi-quasiparticle states and it is now
possible to investigate $\gamma$-bands built on quasiparticle
structures. It is demonstrated in the present work that excited band
structures observed in some nuclei that have remained abstruse for
many years are, in fact, the $\gamma$-bands based on two-quasiparticle
states. In the following, we shall provide results of the TPSM
calculations for $^{132,134}$Ce-, $^{134,136,138}$Nd- and $^{68}$Ge-isotopes where some
excited bands are proposed  as the $\gamma$-bands based on
two-quasiparticle configurations. 
\begin{figure}[thb]
\begin{center}
  \includegraphics[width=0.9\columnwidth]{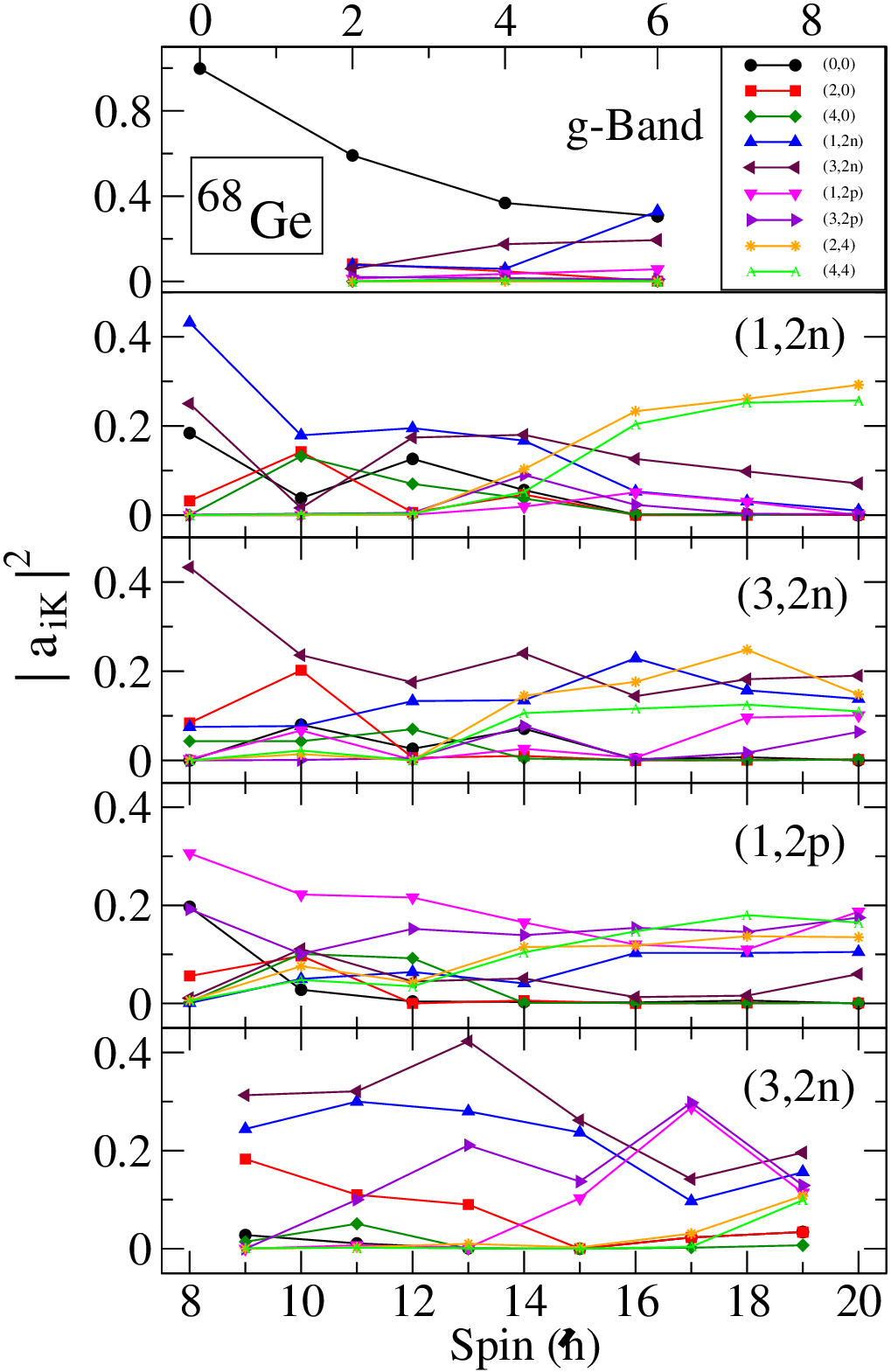}
\end{center}
\caption{\label{fig.1} (Color
online) Probability of various projected K-configurations in the
wavefunctions of the observed  bands for $^{68}$Ge. For clarity, only
the lowest projected K-configurations in the wavefunctions of bands
are shown and in the numerical calculations, projection has been 
performed from more than
forty intrinsic states. 
}\label{wave005}
\end{figure}
 %
\begin{figure}[thb]
\begin{center}
  \includegraphics[width=0.9\columnwidth]{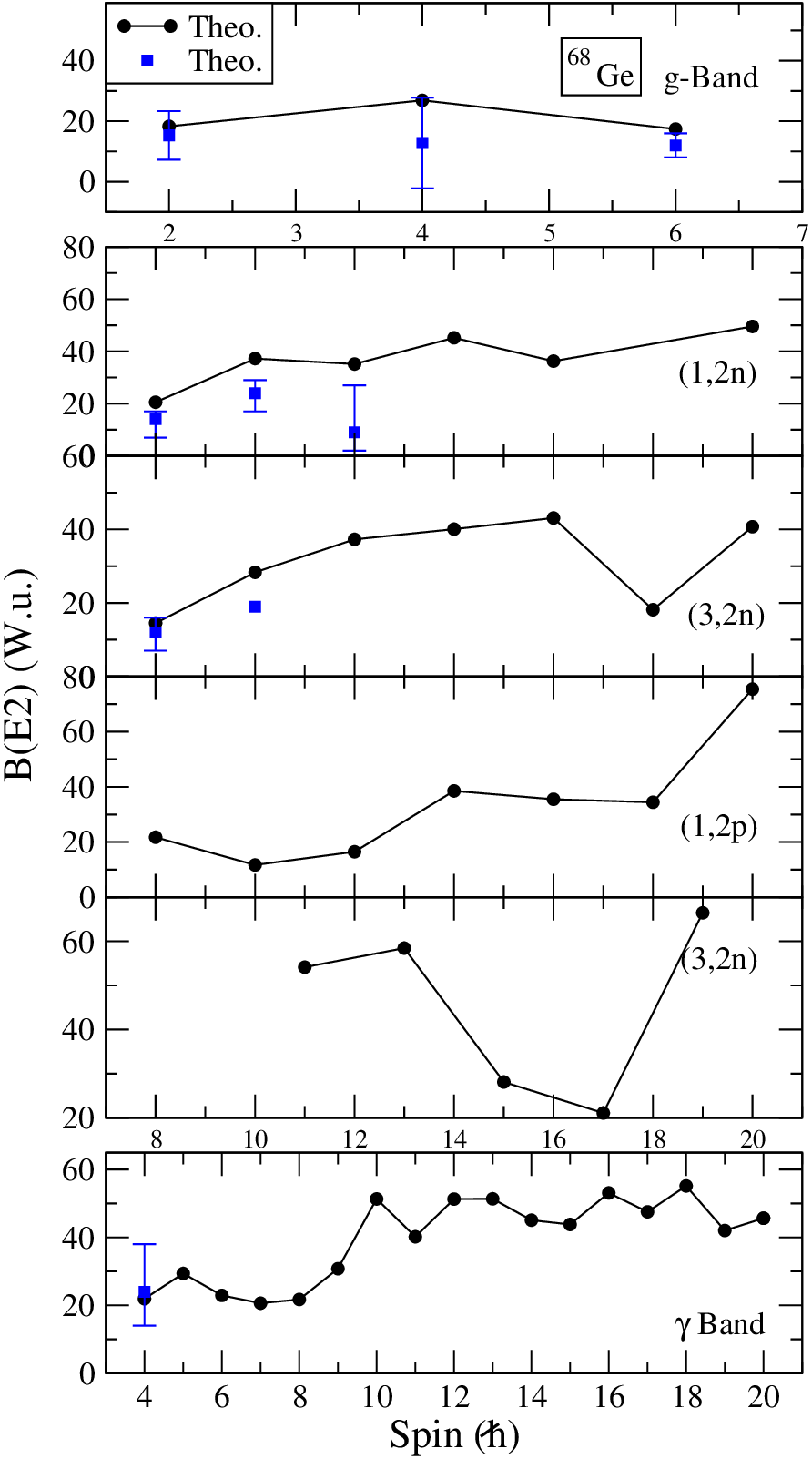}
\end{center}
\caption{\label{fig.1} (Color online) Comparison of the calculated  B(E2) (W.u.) with available experimental data for $^{68}$Ge. Data has been taken from Refs. \cite{75mb}.}\label{figtpsm4}
\end{figure}

In $^{134}$Ce, the ground-state band is observed to fork into two s-bands and the band
heads of both these bands are known to 
have negative g-factors, indicating that both
of them have neutron configurations \cite{Ze82}. In many rotational nuclei, the
ground-state band is crossed by a two quasiparticle state having pair 
of particles with angular-momentum aligned along the rotational axis  \cite{EG73}.
These two-quasiparticle bands  become favoured at some
spin, depending on the region, are referred to as the s-bands. For a
class of nuclei in $A=130$ region, both protons and neutrons 
occupy same aligning (high-j) configuration
with the result that both two-neutron and two-proton states cross 
the ground-state almost simultaneously thus resulting  into the 
forking of the ground-state band \cite{wyss,108Pd}. It is, therefore, expected that one s-band should
have positive g-factor corresponding to the proton configuration
and the other s-band must have negative g-factor as it belongs 
to the neutron configuration. However, observation of negative 
g-factors for both the s-bands is quite puzzling and this problem has remained
unresolved for many years. In the following, the results of TPSM study
are presented that clearly demonstrated that second band is the
$\gamma$-band based on the two-quasineutron states having the 
same intrinsic configuration, and therefore, the two s-bands are
expected to have the similar g-factors.

TPSM study for $^{132,134}$Ce and $^{134,136,138}$Nd isotopes has been
 performed with both neutrons and protons in N = 3, 4 and 5 shells
 and with pairing strength parameters of $G_1=20.82$ and $G_2=13.58$. 
The calculations have been performed
with the deformation parameters displayed in Table \ref{defcd}.
 The band diagrams of $^{134}$Ce
and $^{138}$Nd are provided in Fig.~\ref{fig.ce1} as representative
examples. The band diagram depict the projected energies for different
configurations before diagonalization of the  shell model Hamiltonian
and  are quite instructive as they provide information on the underlying
 intrinsic structures of the bands. The bands in Fig.~\ref{fig.ce1}
 and in
other band diagrams, presented later in this article, are
labeled as  $(K,nqp)$,  where $nqp$ is the number of
 quasiparticles in
a given configuration. For instance, $(0,0)$ is the projection from
the vacuum configuration with K = 0 and corresponds to the ground-state
band. The normal $\gamma$-band with configuration of $(2,0)$ is the 
first excited band and is noted to depict quite large odd-even
staggering for both the systems. For $^{134}$Ce, the ground-state band
is crossed by two-quasineutron configuration, $(1,2n)$, at I=8 and
above this spin value the yrast states originate from this quasiparticle
configuration. 

What is most interesting to note from Fig.~\ref{fig.ce1} is 
that the configuration $(3,2n)$, which is the $\gamma$-band built on
the two-neutron configuration $(1,2n)$, also crosses the ground-state 
band at I=10. It is also evident that two-proton aligned structure,
$(1,2p)$, also crosses the ground-state at I=10, but  is higher in energy
than the configuration $(3,2n)$. Although, the final placement of the
band structures shall vary after considering the configuration mixing,
but it is expected that lowest two s-bands in $^{134}$Ce to emerge from
the same neutron configuration as revealed through the g-factor
measurements. The band diagram for $^{138}$Nd, shown in the lowest
panel of Fig.~\ref{fig.ce1}, depict a completely  different behaviour as
compared to $^{134}$Ce with two two-proton aligned bands crossing the 
ground-state band at I=10. The first two-proton band that crosses
has the configuration $(1,2p)$ and the second has the configuration
$(3,2p)$, which is the $\gamma$-band based on the parent two-proton
configuration. It is, therefore, expected that two s-bands observed in
$^{138}$Nd should both have positive g-factors. 
\begin{table*}
\caption{Axial and triaxial quadrupole deformation parameters
$\epsilon$ and $\epsilon'$  employed in the TPSM calculation. Axial deformations are taken from \cite{Raman}
and nonaxial deformations are chosen in such a way that band heads
of the $\gamma-$ bands are reproduced.}
\begin{tabular}{|ccccccccc| }
\hline        & $^{154}$Dy &$^{158}$Dy     &  $^{160}$Dy  &$^{162}$Dy &$^{164}$Dy & $^{180}$Hf& $^{180}$Hg & $^{238}$U \\
\hline $\epsilon$ & 0.262 & 0.260  &   0.270     & 0.280   & 0.252  & 0.215  &  0.220    &0.210  \\
       $\epsilon'$&0.100     & 0.110   & 0.110   &0.120      & 0.130 & 0.100     & 0.0.90  & 0.085\\\hline
\end{tabular}\label{tab1}
\end{table*}
The obtained s-band structures obtained after diagonalization of the
shell model Hamiltonian, Eq.~\ref{hamham}, are shown in 
Fig.~\ref{fig.ce2} with the 
available experimental data for all  the studied Ce- and Nd-isotopes.
In this figure four s-bands are plotted with dominant components 
from the configurations, $(1,2n), (3,2n), (1,2p)$ and $(3,2p)$. For 
$^{132}$Ce, the lowest two s-bands originate from the configurations,
$(1,2n)$ and $(1,2p)$, which are normal neutron and proton s-bands.
The $\gamma$-bands built on these two-quasiparticle structures 
are quite high in energy. Two observed s-bands for $^{132}$Ce, also
shown in the figure,  are noted to be reproduced quite well by TPSM
calculations and g-factors for the two s-bands are predicted to have
opposite signs as one corresponds to the proton and the other to the
neutron configuration. In the case of $^{134}$Ce, the lowest two
calculated s-bands have configurations of $(1,2n)$ and $(3,2n)$ and,
therefore, both the observed s-bands are predicted to have neutron
configuration as the second s-band is the $\gamma$-band built 
on the two-neutron configuration. As a matter of fact, the g-factor
measurements have been carried out for the band head states, I=10,
for the two s-bands and both have negative g-factors, confirming
that both states are based on the neutron configuration \cite{me86}. 

For $^{134}$Nd, the two s-bands are predicted to have $(1,2p)$ and 
$(1,2n)$ configurations and, therefore, it is predicted that two observed
s-bands for this system should correspond to normal proton
and neutron structures. In $^{138}$Nd, a completely different
scenario is predicted for the lowest two s-bands with both of them
having proton intrinsic structure. The lowest s-band is predicted
to have $(1,2p)$ configuration and the second s-band has $(3,2p)$
configuration which is the $\gamma$-band based on the preceding
configuration. It would be quite interesting to perform the g-factor
measurements for this system as both the s-bands are expected to have
positive values.


\begin{figure}[thb]
\begin{center}
\includegraphics[width=0.9\columnwidth]{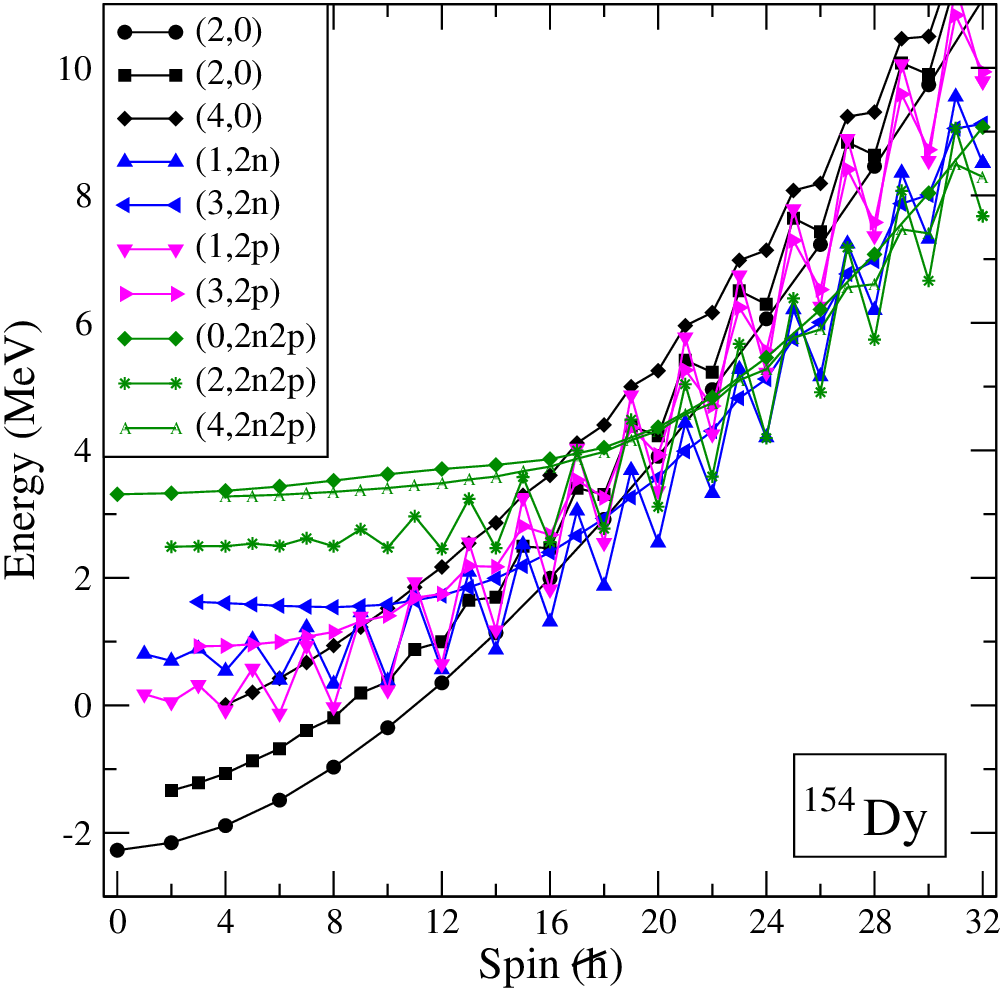}
\end{center}
\caption{(Color online) Band diagram for $^{154}$Dy nucleus. The labels (0,0), (2,0), (4,0),
(1,2n), (3,2n), (1,2p), (3,2p), (0,2n2p)  (2,2n2p) and (4,2n2p) 
correspond to ground,
$\gamma$, 2$\gamma$, two neutron-aligned, $\gamma$-band based
this two neutron-aligned state, two proton-aligned, $\gamma$-band built 
on this two proton-aligned state, two-neutron plus two-proton aligned,  
$\gamma-$ and $\gamma\gamma-$ band built on this four-quasiparticle state.
 }
\label{dyfig1}
\end{figure}
\begin{figure}[thb]
\begin{center}
\includegraphics[width=0.9\columnwidth]{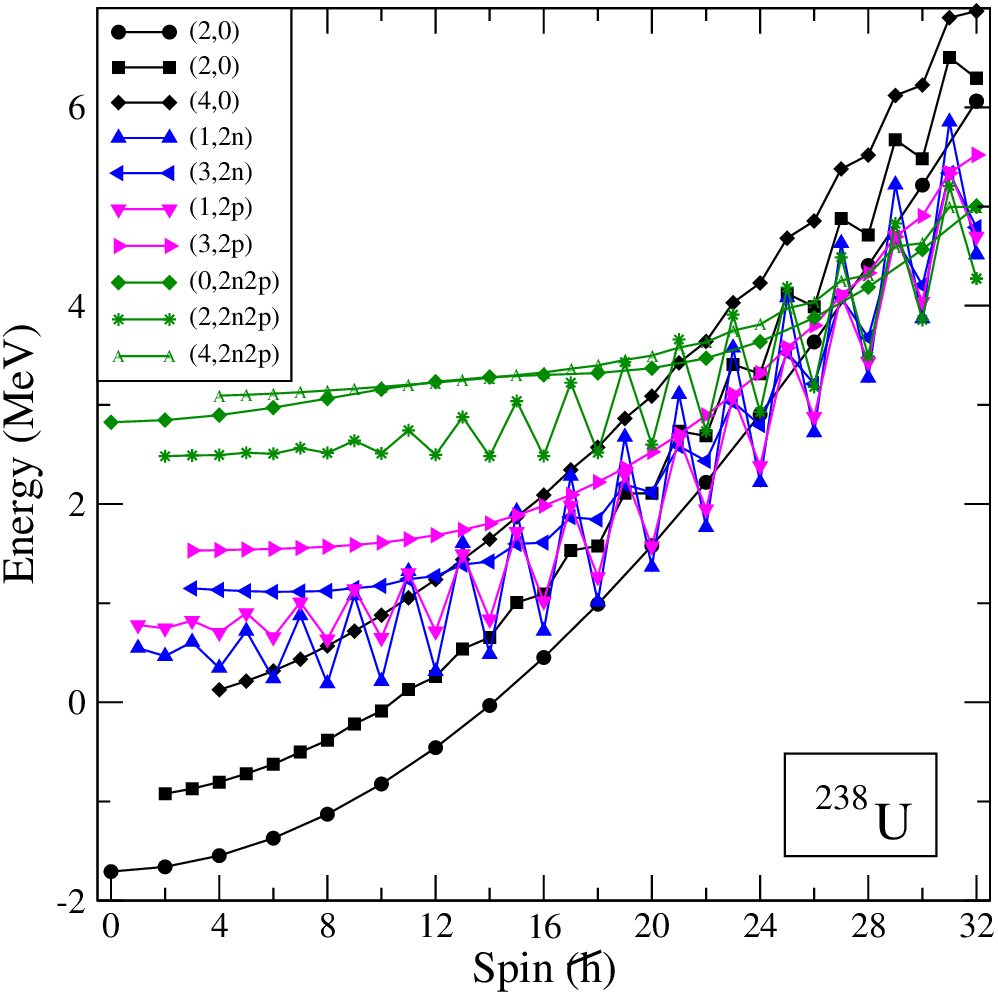}
\end{center}
\caption{(Color online) Band diagram for $^{238}$U nucleus. The labels (0,0), (2,0), (4,0),
(1,2n), (3,2n), (1,2p), (3,2p), (0,2n2p)  (2,2n2p) and (4,2n2p) 
correspond to ground,
$\gamma$, 2$\gamma$, two neutron-aligned, $\gamma$-band on 
this two neutron-aligned state, two proton-aligned, $\gamma$-band on 
two this proton-aligned state, two-neutron plus two-proton aligned,  
$\gamma-$ and $\gamma\gamma-$ band built on this four-quasiparticle state.
 }
\label{dyfig2}
\end{figure}
\begin{figure}[thb]
\begin{center}
\includegraphics[width=0.90\columnwidth]{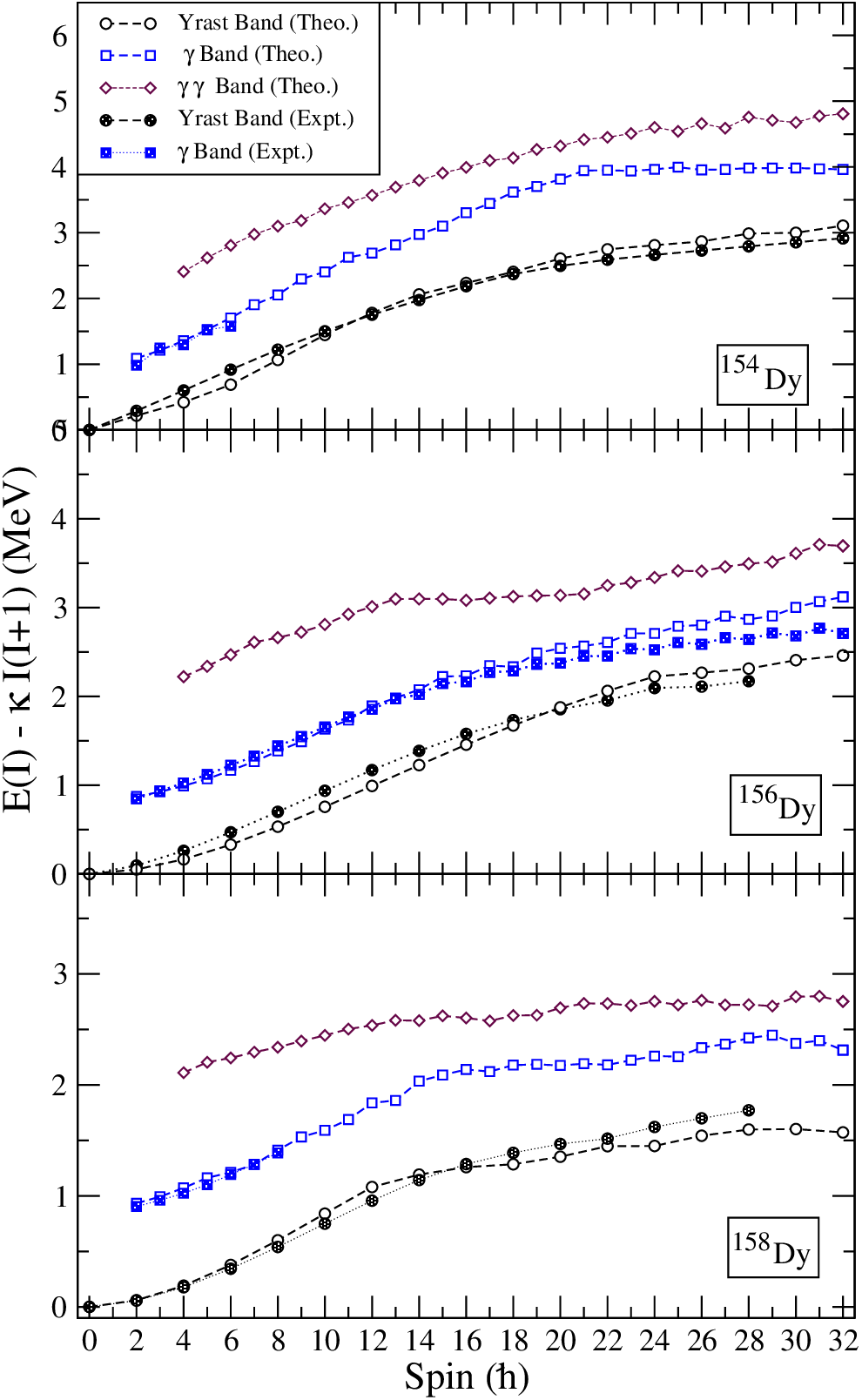}
\end{center}
\caption{\label{fig.1} (Color online) Comparison of the measured energy levels of
yrast-, $\gamma$-, and $\gamma$$\gamma$-bands for $^{154,156,158}$Dy
 nuclei and the results of TPSM calculations. The
scaling factor $\kappa$ appearing in the y-axis is defined as
$\kappa=32.32 A^{-5/3}$. Data is taken from
Refs.~\cite{154dy,156dy,158dy}.}\label{dyfig3}
\end{figure}
\begin{figure}[thb]
\begin{center}
\includegraphics[width=0.90\columnwidth]{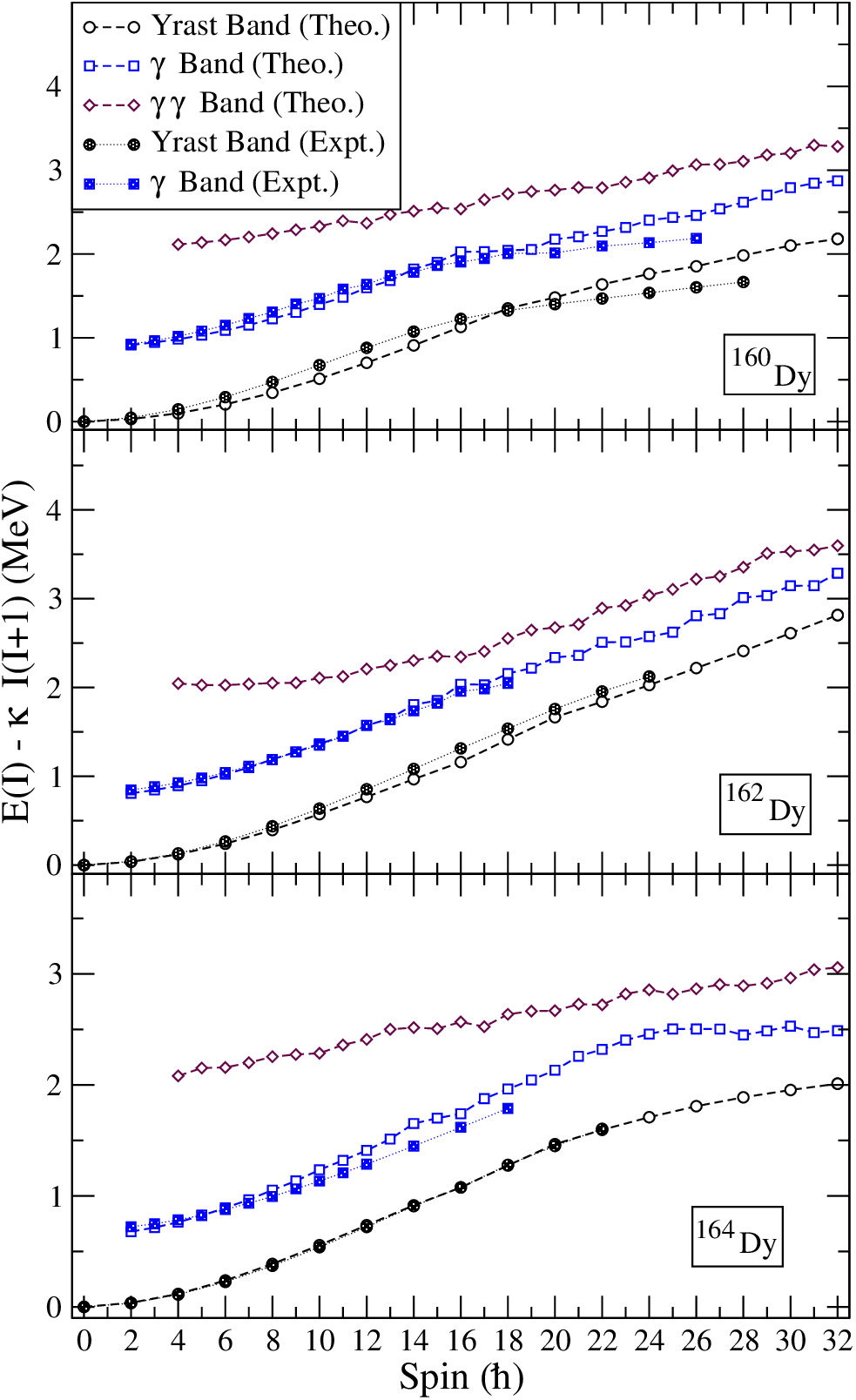}
\end{center}
\caption{\label{fig.1} (Color online) Comparison of the measured energy levels of
yrast-, $\gamma$-, and $\gamma$$\gamma$-bands for $^{160,162,164}$Dy
 nuclei and the results of TPSM calculations. The
scaling factor $\kappa$ appearing in the y-axis is defined as
$\kappa=32.32 A^{-5/3}$. Data is taken from
Refs.~\cite{160dy,162dy,162dya,164dy}.}\label{dyfig4}
\end{figure}
\begin{figure}[thb]
\begin{center}
\includegraphics[width=0.90\columnwidth]{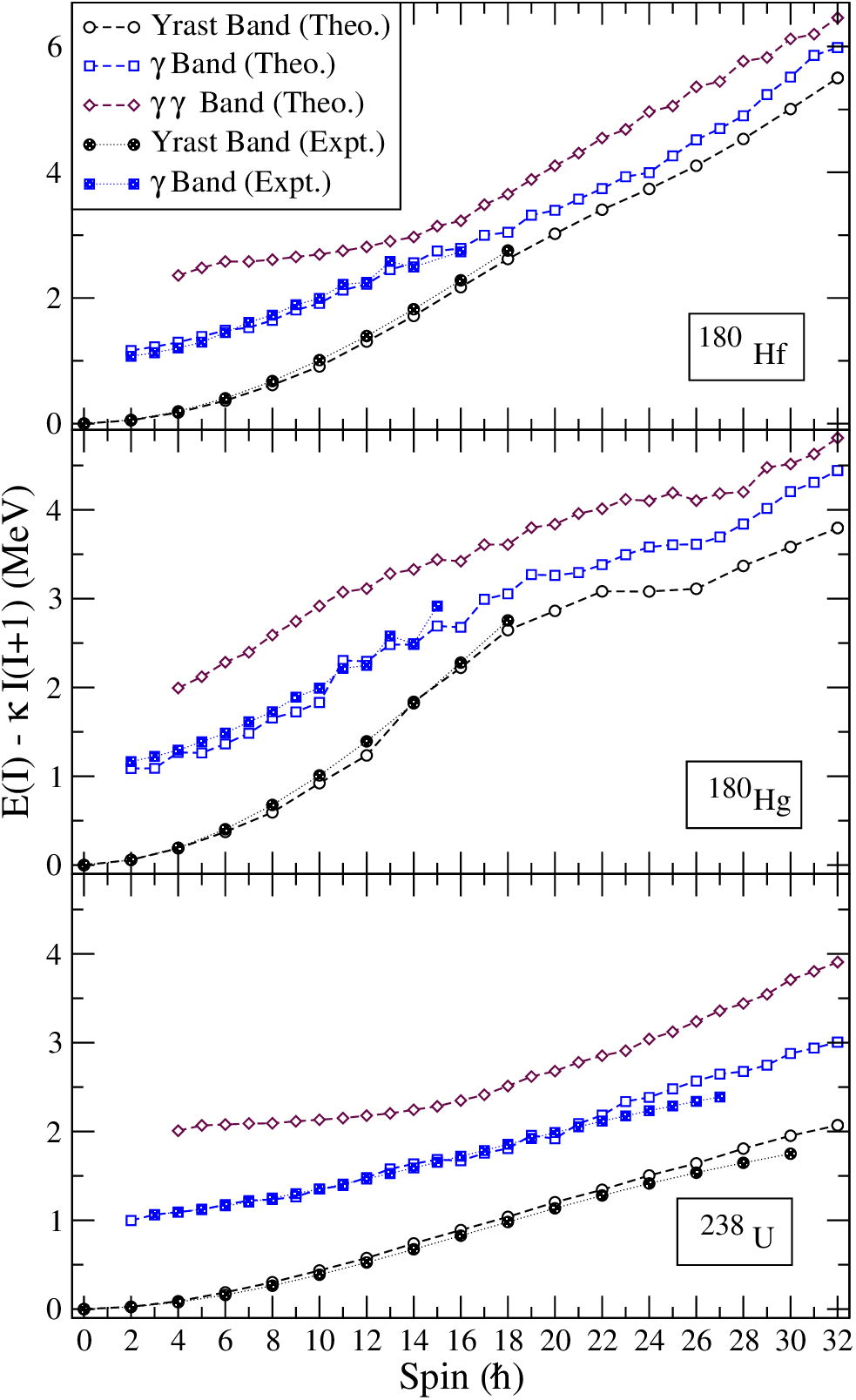}
\end{center}
\caption{\label{fig.1} (Color online) Comparison of the measured energy levels of
yrast-, $\gamma$-, and $\gamma$$\gamma$-bands for $^{180}$Hf, $^{180}$Hg and $^{238}$U 
 nuclei and the results of TPSM calculations. The
scaling factor $\kappa$ appearing in the y-axis is defined as
$\kappa=32.32 A^{-5/3}$. Data is taken from
Refs.~\cite{180hf,180hg,238u}.\label{dyfig5}
}
\end{figure}

\begin{figure}[htb]
\begin{center}
\includegraphics[width=1.0\columnwidth]{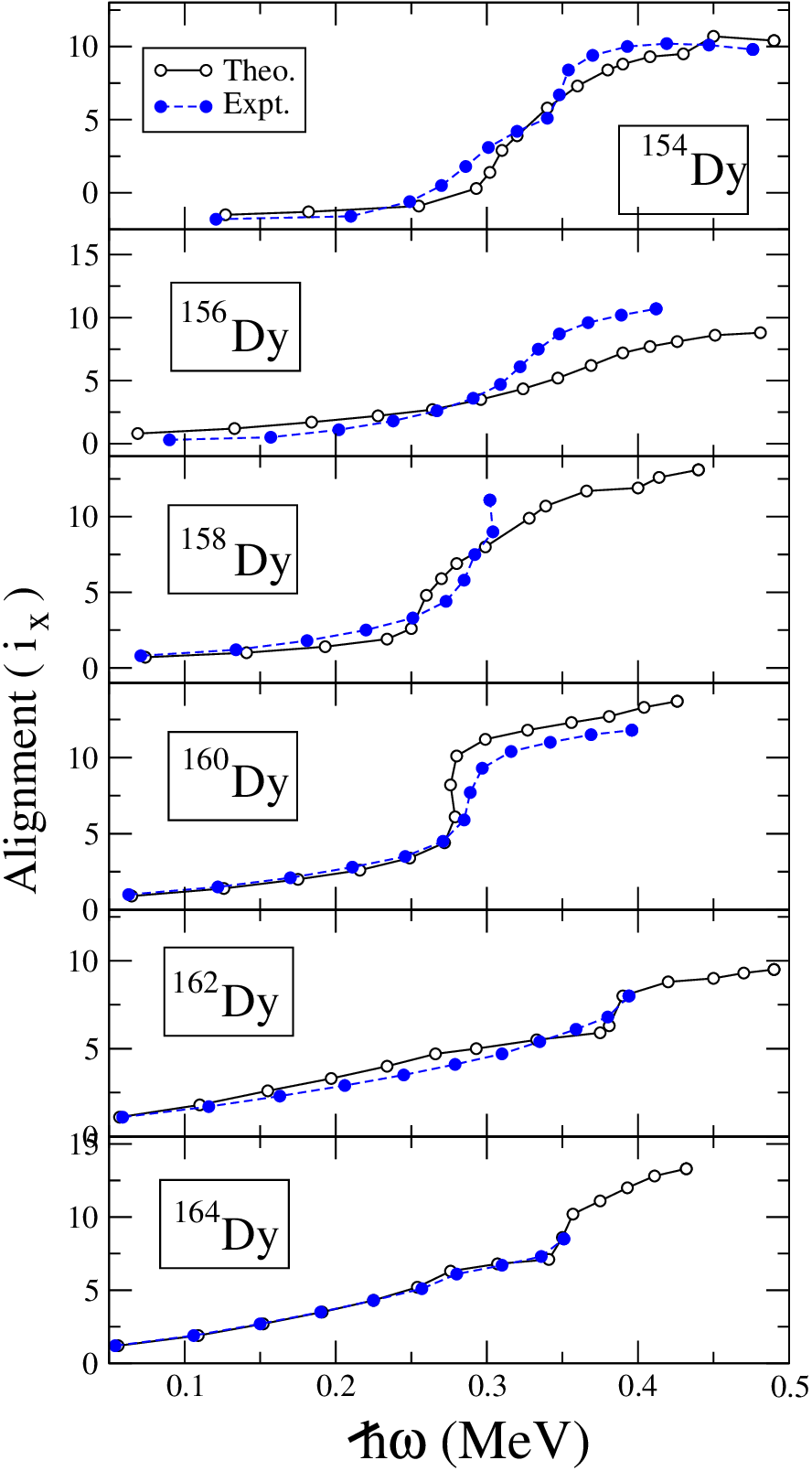}
\end{center}
\caption{\label{fig.1}( Color online) Comparison of the aligned angular momentum $i_x$)
 as a function of frequency, obtained from the measured energy levels
 as well as those calculated
from the TPSM results, for  $^{154-164}$Dy nuclei.
}\label{dyfig6}
\end{figure}

\begin{figure}[htb]
\begin{center}
\includegraphics[width=1.0\columnwidth]{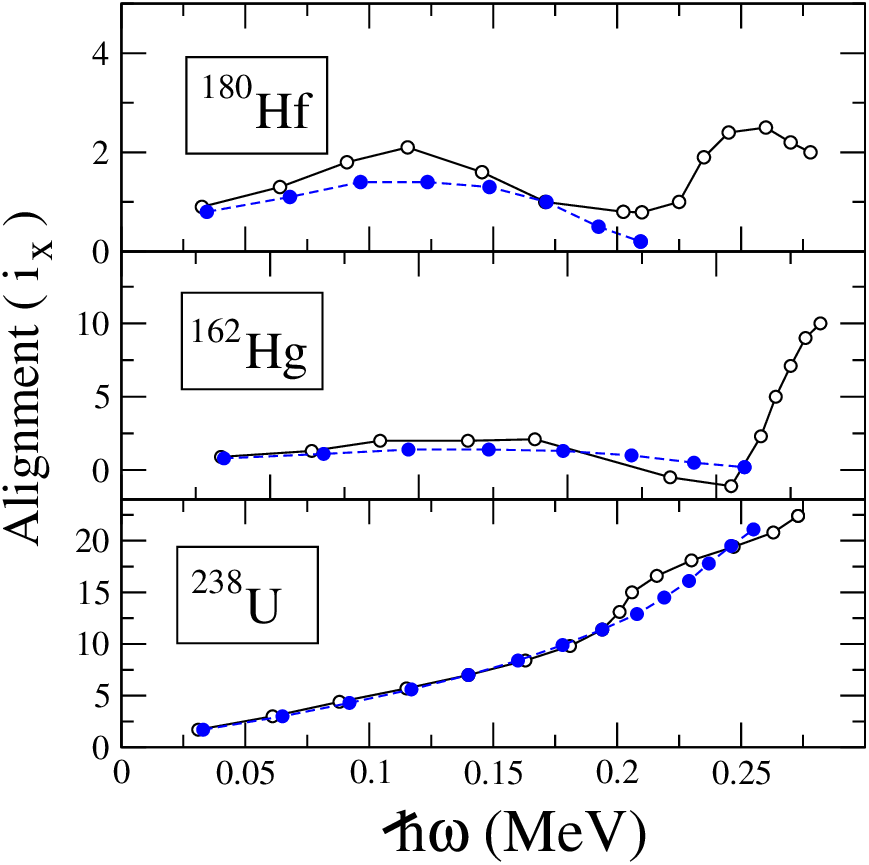}
\end{center}
\caption{\label{fig.1}( Color online) Comparison of the aligned angular momentum $i_x$)
 as a function of frequency, obtained from the measured energy levels
 as well as those calculated
from the TPSM results, for  $^{180}$Hf, $^{180}$Hg and $^{238}$U nuclei.
}\label{dyfig7}
\end{figure}

\begin{figure}[htb]
 \centerline{\includegraphics[trim=0cm 0cm 0cm
0cm,width=0.5\textwidth,clip]{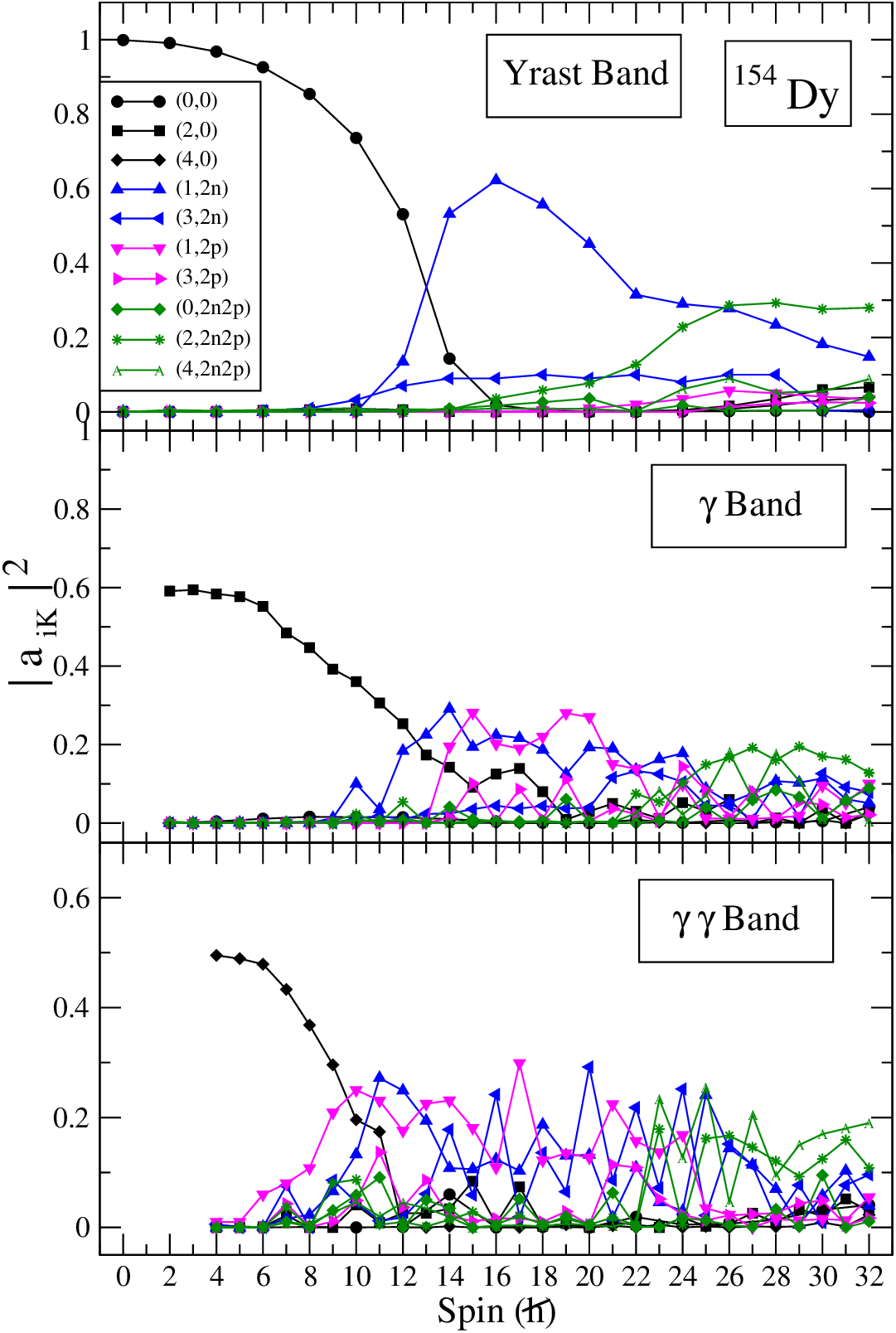}}
\caption{(Color online) Probability of various projected K-configurations in the wavefunctions
of the yrast-, $\gamma-$ and $\gamma\gamma-$bands in $^{154}$Dy.}
\label{dyfig8}
\end{figure}
\begin{figure}[htb]
 \centerline{\includegraphics[trim=0cm 0cm 0cm
0cm,width=0.5\textwidth,clip]{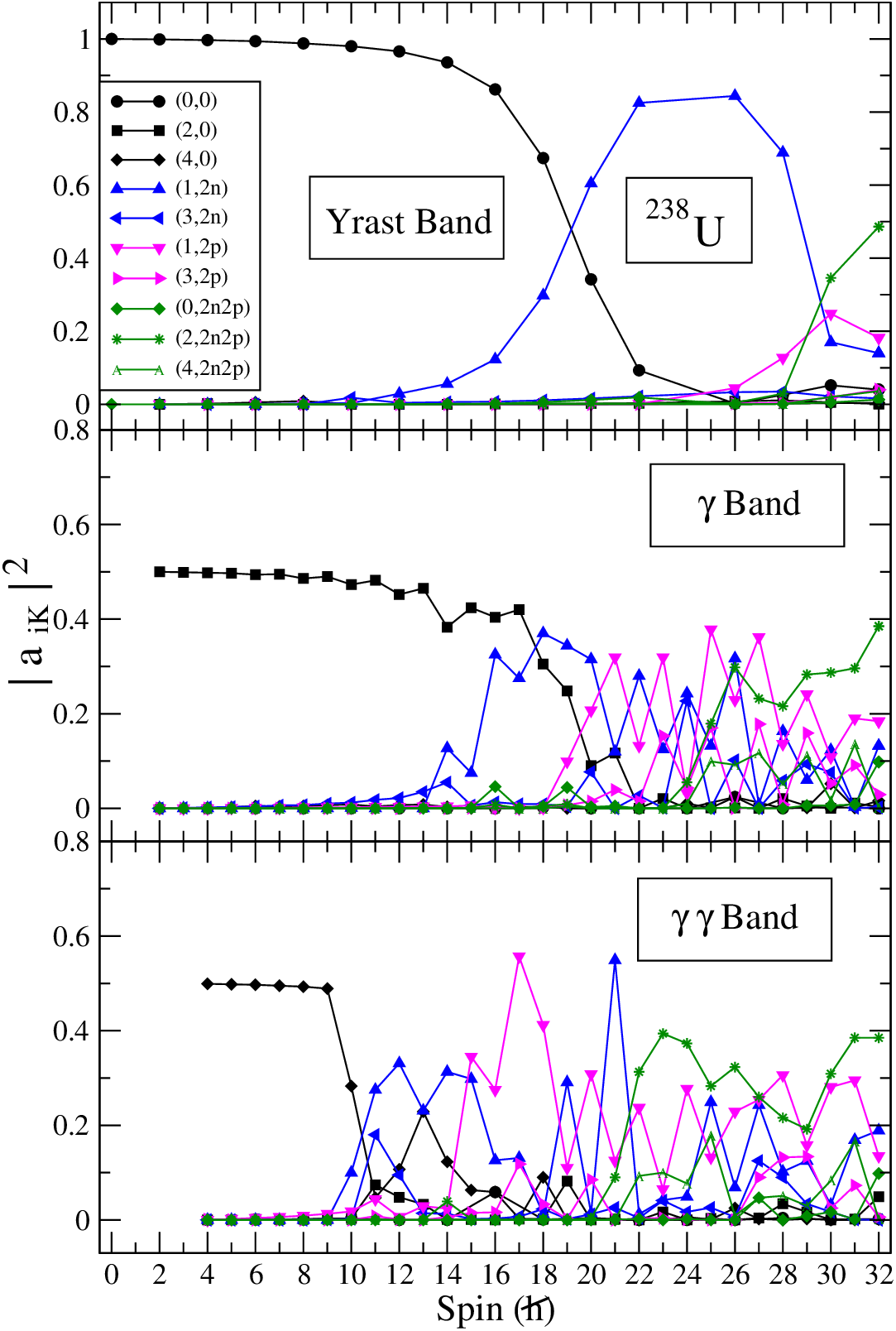}}
\caption{(Color online) Probability of various projected K-configurations in the wavefunctions
of the yrast-, $\gamma-$ and $\gamma\gamma-$bands in $^{238}$U.}
\label{dyfig9}
\end{figure}
\begin{figure}[thb]
\begin{center}
\includegraphics[width=0.9\columnwidth]{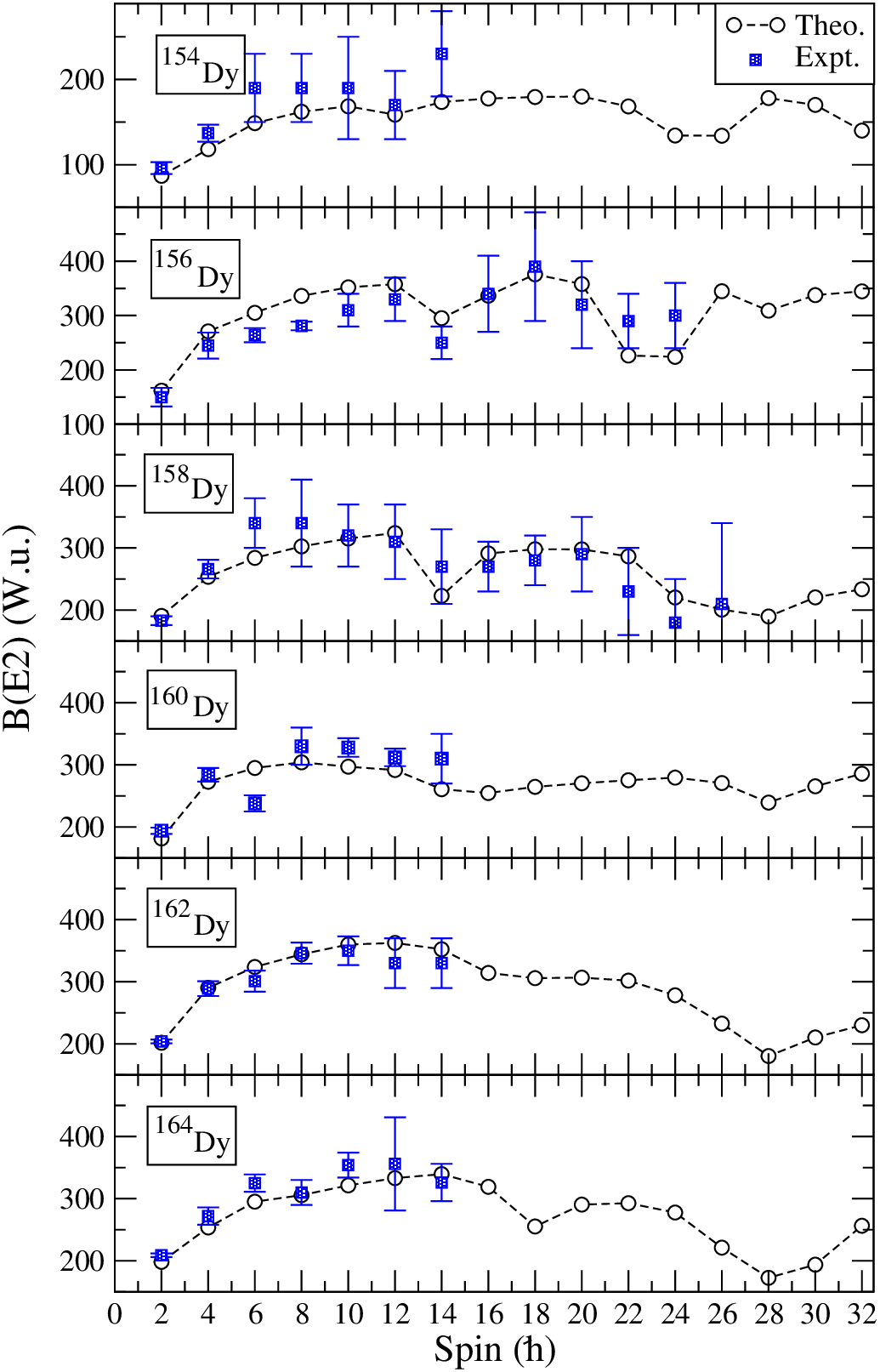}
\end{center}
\caption{\label{fig.1} (Color online) Comparison of experimental and calculated $B(E2)$ for $^{
154-164}$Dy. Data is taken from
Refs.~\cite{154dy,156dynd,158dynd,160dynd,162dynd,164dynd}.
}\label{dyfig10}
\end{figure}
\begin{figure}[thb]
\begin{center}
\includegraphics[width=0.9\columnwidth]{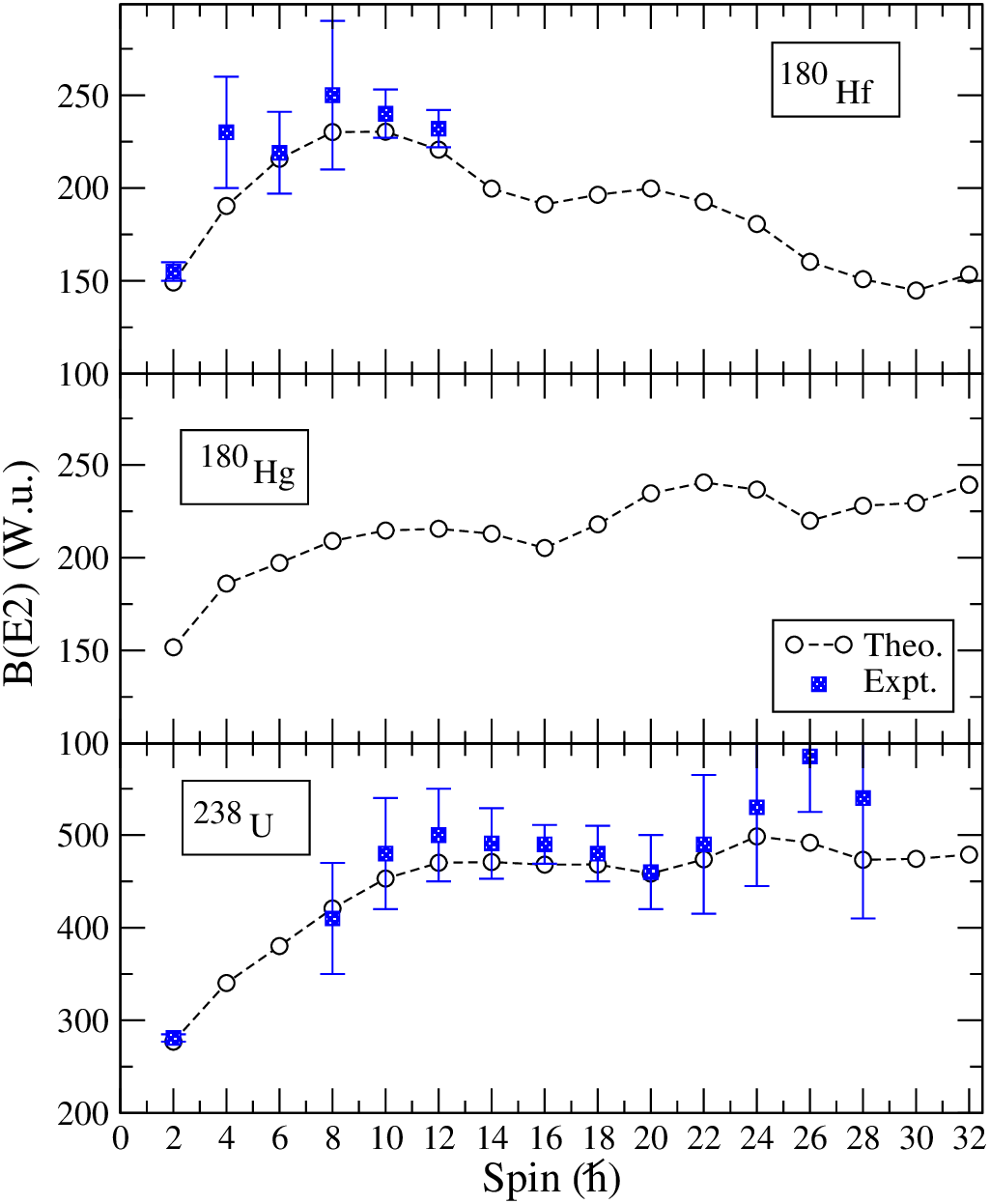}
\end{center}
\caption{\label{fig.1} (Color online) Comparison of experimental and calculated $B(E2)$ for $^{
180}$Hf, $^{180}$Hg and $^{238}$U . Data is taken from
Refs.~\cite{180hg,238unds}.}\label{dyfig11}
\end{figure}

It is expected that $\gamma$-bands based on the multi-quasiparticle
states should be more wide spread in nuclear periodic table. In the
following we shall report a recent study for $^{68}$Ge where multiple
s-bands are also observed and it is shown that one of the observed s-band
in $^{68}$Ge has the structure of the $\gamma$-band built on the two-quasineutron 
configuration. TPSM calculations for $^{68}$Ge have been performed with
the deformation parameters, $\epsilon = 0.22$ and $\epsilon'=0.16$
using the oscillator space of $N = 3, 4$ and $5$ 
(both for protons and neutrons) and 
the pairing strengths of $G_1=20.82$ and $G_2=13.58$. 
 Fig.~\ref{figtpsm1} is the obtained band
diagram for $^{68}$Ge and it is seen that $\gamma$-band depicts quite
large odd-even staggering with the even-spin members closely following
the ground-state band. It is further noted that ground-state band is
crossed by two-quasiparticle band having $(1,2n)$ and $(3,2n)$
configurations at I = 8. These
two bands are two-neutron aligned and the $\gamma$-bands built on this
two-neutron state. The odd-spin members from the two-neutron
$\gamma$-band form the yrast states from I = 9 to 13. It is also evident
from Fig.~\ref{figtpsm1} that two-aligned band having $(1,2p)$ configuration also
crosses the ground-state band at I = 10 and from I = 14 the
four-quasiparticle formed from two-neutron plus two-proton
configuration become lowest in energy. 

The band structures obtained after diagonalization are displayed in
Fig.~\ref{figtpsm2} along with the experimental data which depicts multiple
s-bands above the ground-state band. It is evident from the figure
that TPSM approach reproduces the observed band structures remarkably 
well. TPSM also predicts many new states for the $\gamma$-band and
also a few lower states for the bands labelled as $(1,2p)$ and
odd-spin member of $(3,2n)$. These levels have been assigned based
on the dominant components in the calculated wavefunctions plotted 
in Fig.~\ref{wave005}. The ground-state band, shown in the top panel of
 Fig.~\ref{wave005},
has the dominant component, as expected, of $(0,0)$ for I = 0, 2 and
4. For I = 6, the contribution of the two-neutron aligned state having
$(1,2n)$ configuration becomes equally important. It is noted from
the wavefunction analysis that two observed s-bands beginning with I = 8
have dominant components of $(1,2n)$ and $(3,2n)$ configurations
and, therefore, both these bands have neutron configuration and one
of them is the $\gamma$-band based on the two-quasineutron 
configuration. The third observed s-band beginning with I = 12 has
the $(1,2p)$ configuration. It needs to be clarified that it is somewhat erraneous
to label these bands with a specific two-quasiparticle configuration as
there is a substantial mixing and further for high-spin states 
four-quasiparticle states become lower in energy. 

It is quite clear from above discussion that the lowest two observed
s-bands have both neutron structure and the third s-band has the
proton structure. The g-factor measurements have been performed for 
the lowest two s-bands and  both are confirmed  to have neutron
configuration and, therefore,  validating the TPSM predictions \cite{me86} 
as the
second s-band is the $\gamma$-band built on first s-band and both
originate from their same intrinsic configuration. 

We have also evaluated the intra-band BE2 transition probabilities for
various band structures discussed above and are plotted in Fig.~\ref{figtpsm4}.
Measured BE2 values are also available for some transitions
have been depicted in the figure. For the g-band, the
experimental values are well reproduced by the TPSM calculations,
however, disagreement is clearly seen for the I=10 transition in the two-
quasineutron band. More experimental data is needed to test
the predictions of the TPSM in detail. 

\subsection{Systematic investigation of $\gamma$-band structures in Dy, Hf, Hg and U nuclei}
\label{Sect.04}

A systematic investigation of the band structures
observed in $^{154-164}$Dy, $^{180}$Hf, $^{180}$Hg and $^{238}$U
nuclei have been performed.
 These nuclei have been chosen in the present study as
$\gamma$-bands are known up to highest spin and it is possible to test
the predictions of the TPSM approach in the limit of largest angular
momemtum. TPSM calculations for these nuclei have been performed with
the deformations listed in Table \ref{tab1} in the shell model space
with N $=$ 4, 5, 6 for neutrons and N $=$ 3, 4, 5 for protons
 and the pairing strength parameters of $G_1=21.24$ and
$G_2=13.86$. As representative exmples,
the band diagrams for $^{154}$Dy and $^{238}$U are displayed in
Figs.~\ref{dyfig1} and \ref{dyfig2}.
 In the band diagram, the ground-state band having $(0,0)$
configuration is crossed at I=14 by two-neutron aligned state
configuration, $(1,2n)$. Further, it is noted that four-quasiparticle
configuration, $(2,2n2p)$, crosses the two-neutron state at I=26
and above this spin value, it is expected that four-quasiparticle 
configurations will dominate the yrast band. 

The band diagram for $^{238}$U is displayed in Fig.~\ref{dyfig2}. 
The TPSM calculations for this system were carried out within the
space of N $=$ 5, 6, 7 for neutrons and N $=$  4, 5, 6
 for protons and with the pairing strengths of $G_1=16.80$ and
$G_2=12.80$. It is noted from the figure that the first crossing due to
alignment of two-neutrons occurs at I = 20 and the second due to the
alignment of two-neutrons plus two-protons occurs at I=30. The
two-proton aligned band is seen to remain always
higher in energy as compared to the neutron-aligned
configuration. 

The projected energies for yrast-, $\gamma$- and
$\gamma\gamma$- bands, obtained after diagonalization of the
shell model Hamiltonian, are plotted in Figs. ~\ref{dyfig3}, \ref{dyfig4} and 
\ref{dyfig5}
along with the available experimental data. To have a better comparison
between theoretical and experimental energies, these quantities
  have
been subtracted by a core contribution. It is evident from the
three figures that observed yrast bands in all the
studied nuclei are reproduced fairly well by the TPSM calculations.
For the $\gamma$-bands, the agreement between observed and calculated
energies is also quite good, except that some deviations are noted in
$^{156}$Dy and $^{160}$Dy
 for high-spin states above I = 22. These deviations are also
noted in some of the yrast bands and are expected since for spin above
$22\hbar$ the contributions from  negleced four-quasineutron and
four-quasiproton configurations are anticipated to become important. 

$\gamma\gamma$-bands have not been observed in any of the studied
nuclei and it is noted from Figs. ~\ref{dyfig3}, \ref{dyfig4} and \ref{dyfig5}
 that predicted
band heads of these bands are quite high in energy (more than 2 MeV),
but it is noted that for high-spin states these bands become quite
close to the $\gamma$-bands and it should be possible to populate
them at higher spin. 

To probe the crossing features in the studied nuclei, the alignments 
of the $^{154-164}$Dy,$^{180}$Hf, $^{162}$Hg and
$^{238}$U nuclei are displayed in Figs. ~\ref{dyfig6} and \ref{dyfig7} as a
function of the rotational frequency. In the experimental alignment
plot of $^{152}$Dy, two upbends are clearly observed whereas in the
theoretical plot only a broad upbend is noted. These upbends are due to
the crossing of the aligned quasiparticle configurations along the
yrast line. In order to investigate the structural changes as a
function of angular momentum, the wavefunctions are plotted in 
Fig. ~\ref{dyfig8}. In the top panel of this figure, the yrast
wavefunction depicts the first crossing at I = 14
as due to the alignment of two-quasineutrons, $(1,2n)$, and the second
crossing
at I=18 arises from the alignment of four-quasiparticle
configuration, $(2,2n2p)$. The broad upbend noted in the alignment
figure is due to combination of these crossings. In the experimental
plot, the two crossings are clearly evident and this indicates that
interaction among the crossing bands is overpredicted by the TPSM
calculations. Fig. ~\ref{dyfig8} also depicts the wavefunctions for the
$\gamma$- and $\gamma\gamma$-bands and $\gamma$-band
up to I = 12 has the expected dominat component of $(2,0)$. However,
above this spin value, $\gamma$-band is a mixture of many different 
configurations and ceases to be called as the $\gamma$-band.
$\gamma\gamma$-band has the expected structure of $(4,0)$
configuration up to I = 9 and above this spin value it ceases to 
be called as $\gamma\gamma$-band as well.

TPSM calculated  alignment for $^{156}$Dy, shown in the second panel
of Fig.~\ref{dyfig6}, again shows smoother behaviour as compared to the
experimental alignment and is an indication that interaction among the
bands is overestimated in the calculated alignment. For $^{154-164}$Dy,
shown in Fig.~\ref{dyfig6}, and for $^{180}$Hf, $^{162}$Hg and $^{238}$U,
depicted in Fig.~\ref{dyfig7}, the agreement between the calculated
and the experimental values is better than the  previous two cases. 

The wavefunctions for the yrast-, $\gamma-$ and $\gamma\gamma$-bands
are displayed in Fig.~\ref{dyfig9} for $^{238}$U. The cross over between the
ground-state configuration and the two-neutron aligned band is 
noted at I = 20 for the yrast band. Above I=26, two-proton aligned
configuration is also noted to become important and above I=30, the
four-quasipartcle configuration becomes dominant. For the
$\gamma$-band, $(0,2)$ is the dominant configuration up to I=16, but
above this spin value it is a highly mixed band. For
$\gamma\gamma$-band, $(0,4)$, is the dominant configuration up to
I=10 and above this spin, it is again a highly mixed band.

The measured BE2 transition probabilities are available along the yrast band for most of the
nuclei  studied in this section and are depicted in 
Figs.~\ref{dyfig10} and \ref{dyfig11} along
with the calculated BE2 transitions 
using the TPSM wavefunctions and the projected expression given in
Eq. \ref{beeee2}. It is evident from the figure that calculated BE2 reproduce
the known transitions remarkably well, in particular, the drops in the 
measured transitions for $^{156}$Dy, $^{158}$Dy and $^{238}$U. The
first  drop in the transitions are related to the crossing of
two-quasineutron aligned band with the ground-state band and the
second drop is due to the crossing of four-quasiparticle composed of
two-quasineutron and two-quasiproton aligned configuration. TPSM
results also depict two drops for $^{156}$Dy, although, not very pronounced
and are due to  two crossings evident from the wavefunction
analysis. These crossings are not apparent in the
alignment plot of Fig.~\ref{dyfig6}, but are noted in the BE2 transitions as
these are more senstive to the structural changes as compared to the
 quantities derived from the energies. 

\subsection{Chiral symmetry and doublet band structures}
\label{Sect.04}

\begin{figure}[thb]
\begin{center}
\includegraphics[width=0.9\columnwidth]{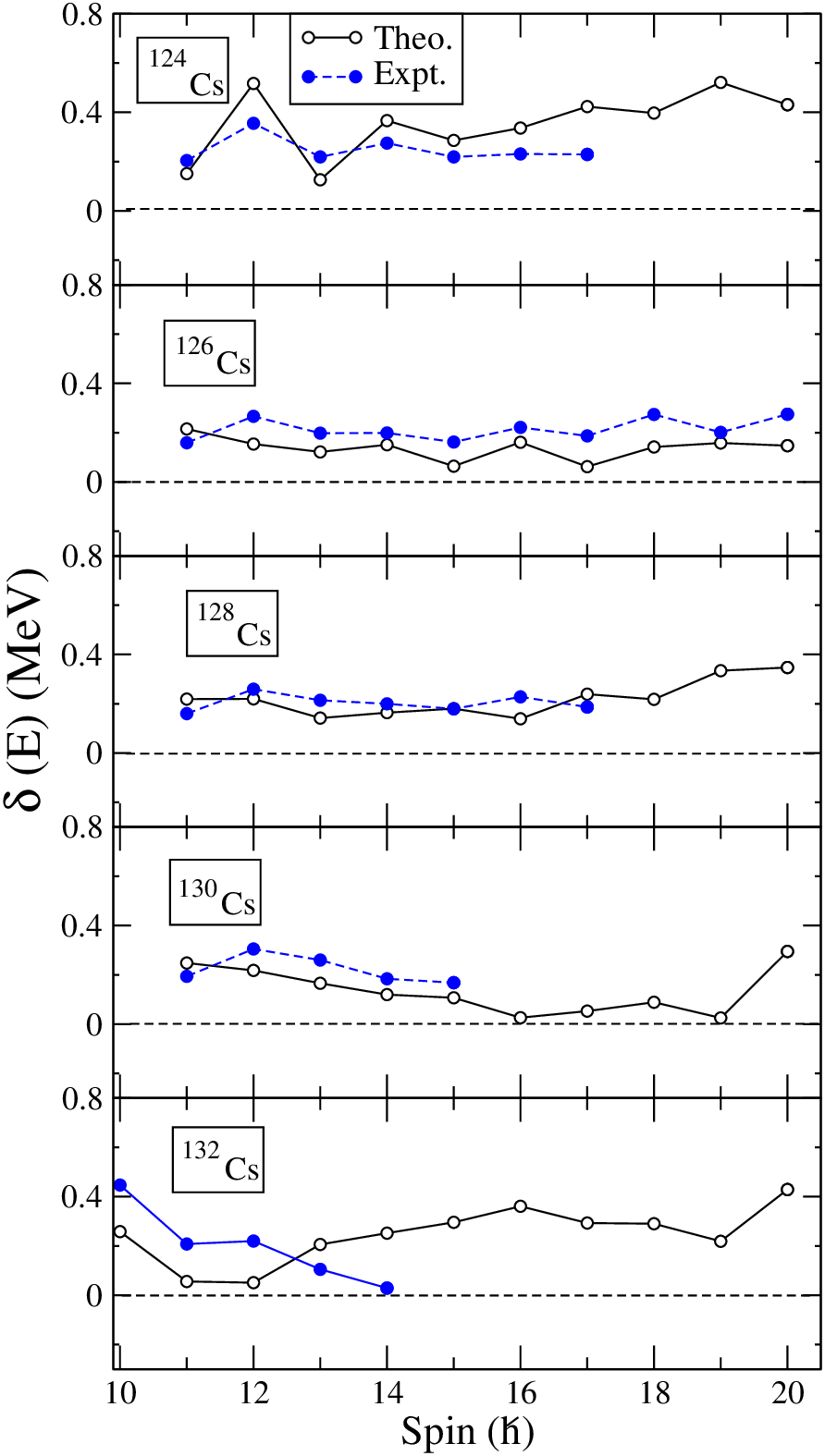}
\end{center}
\caption{ (Color online)
Energy difference, $\delta (E)$, between doublet bands for a given spin. Data has been taken from Refs. 
\cite{EG12,AG01,HG10,EG06,GR03,Simons1}.
} \label{chifig1}
\end{figure}
\begin{figure}[thb]
\begin{center}
\includegraphics[width=0.9\columnwidth]{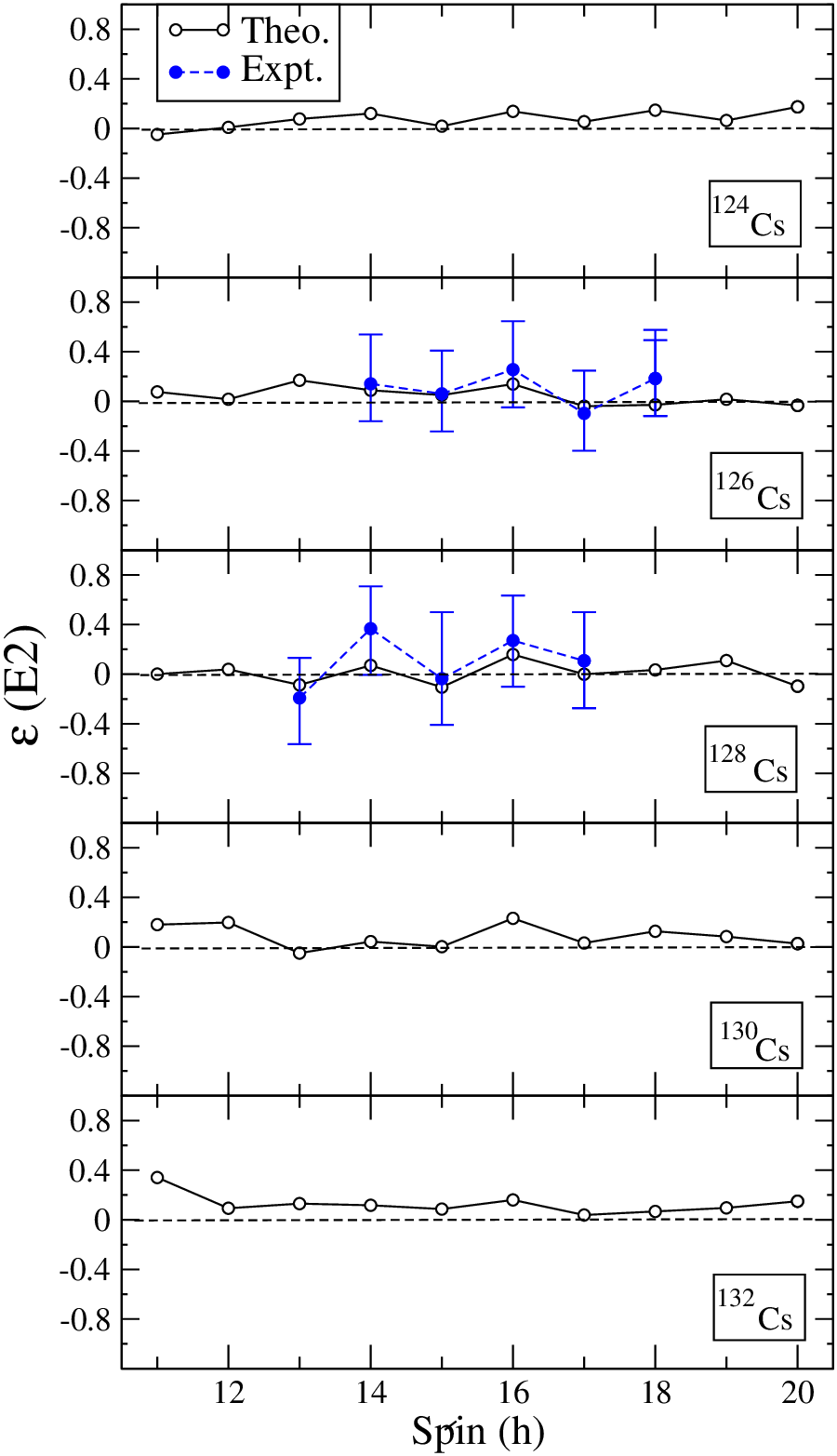}
\end{center}
\caption{ (Color online)
Comparison of the TPSM and  experimental values of  $\epsilon$(E2) as
a function of spin. Data has been taken from Refs. 
\cite{EG12,AG01,HG10,EG06,GR03,Simons1}.
} \label{chifig2}
\end{figure}
\begin{figure}[thb]
\begin{center}
\includegraphics[width=0.9\columnwidth]{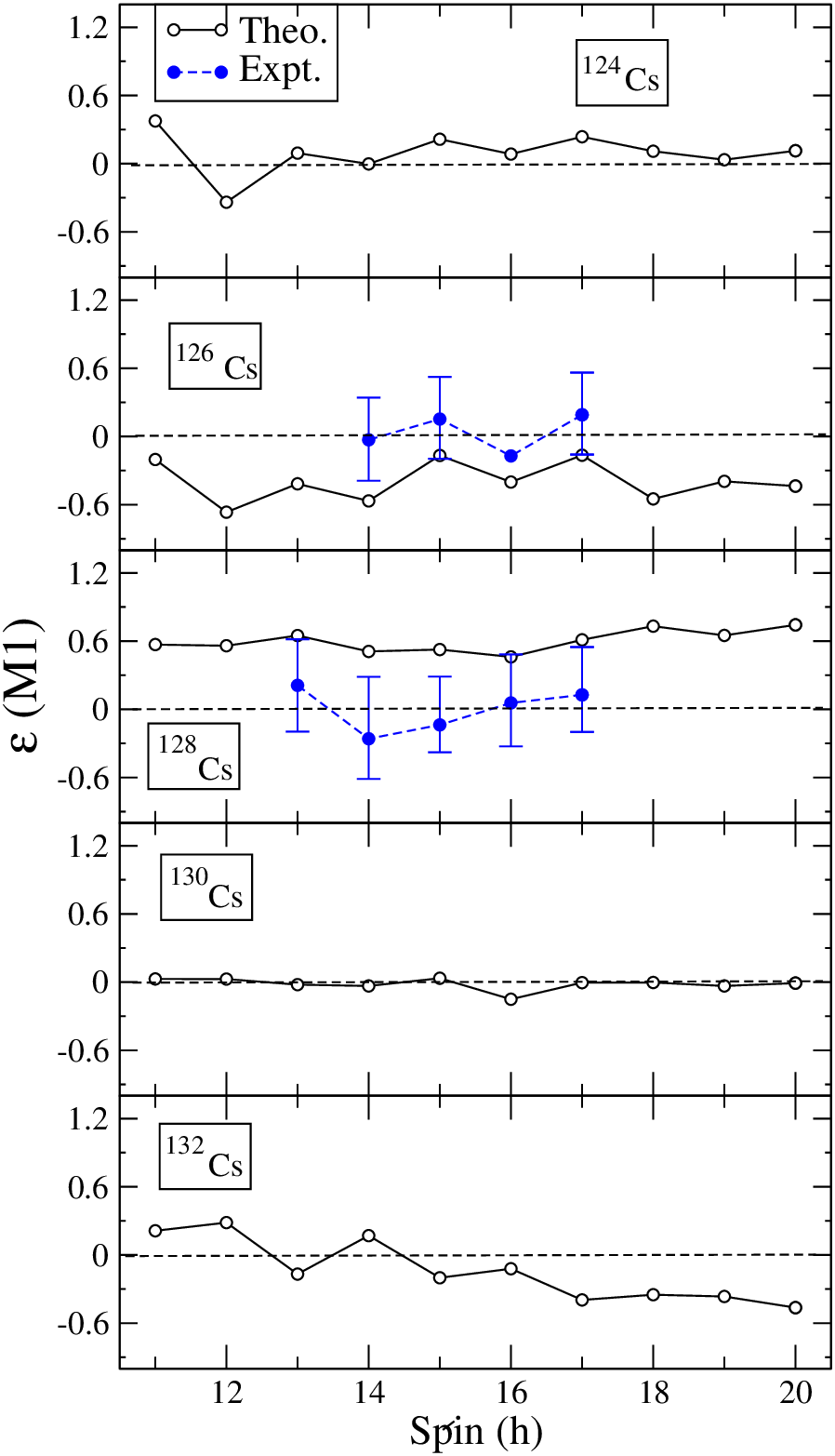}
\end{center}
\caption{\label{fig.1} (Color online)
Comparison of the TPSM and  experimental values of  $\epsilon$(M1) as
a function of spin. Data has been taken from Refs. 
\cite{EG12,AG01,HG10,EG06,GR03,Simons1}.
} \label{chifig3}
\end{figure}

The spontaneous symmetry breaking mechanism has played a
central role in elucidating the intrinsic structures of quantum many
body system \cite{SF01}. What all is possible from the experimental analysis of a
bound quntum many-body system is a set of energy levels that are labelled by
the quantum numbers related to the symmetries preserved by the system.
For instance, for the case of atomic nucleus the energy levels are labelled
by angular-momentum and parity quantum numbers that are related to
the rotational and reflection symmetries. It is through breaking of
these symmetries that provides an insight into the excitation modes of
the atomic nucleus. The most celebrated model that employs the
symmetry breaking mechanism is the Nilsson model. This model breaks
the rotational symmetry and has provided invaluable information on the
structures of deformed nuclei \cite{nilson}. The observation of the rotational band
structures built on each intrinsic state  is a manifestation of the breaking of this
symmetry  \cite{SF97,TS04}.
\begin{figure}[htb]
 \centerline{\includegraphics[trim=0cm 0cm 0cm
0cm,width=0.45\textwidth,clip]{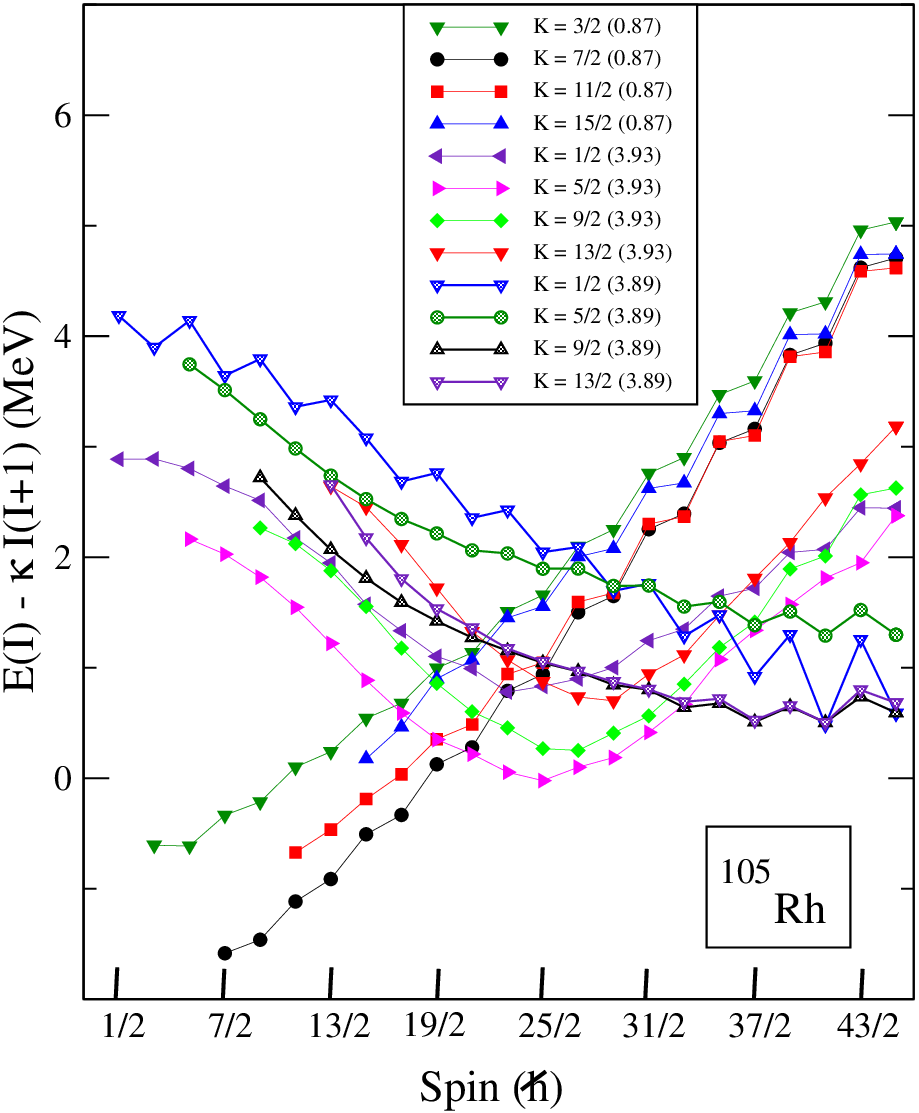}}
\caption{(Color online)  The angular-momentum projected bands obtained
 for different intrinsic K-configuration, given in legend box, for $^{105}$Rh isotopes. The energies 
of the quasiparticle states are given in the parenthesis. The
scaling factor $\kappa$ appearing in the y-axis is defined as
$\kappa=32.32 A^{-5/3}$.}
\label{rhfig1}
\end{figure}

\begin{figure}[htb]
 \centerline{\includegraphics[trim=0cm 0cm 0cm
0cm,width=0.45\textwidth,clip]{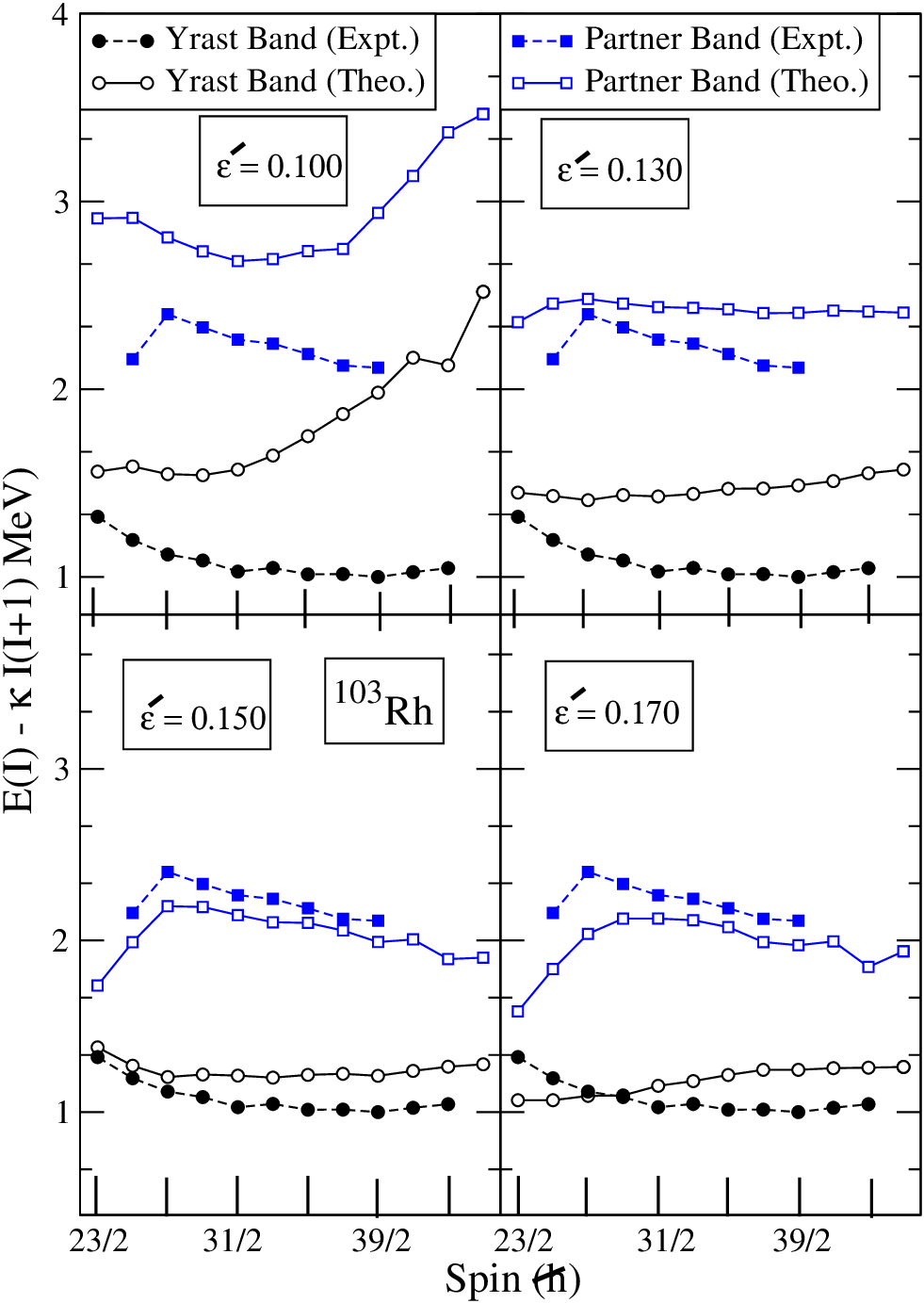}}
\caption{(Color online)  Comparison of the measured energy levels of
of  $^{103}$Rh
 nucleus and the results of TPSM calculations. The
scaling factor $\kappa$ appearing in the y-axis is defined as
$\kappa=32.32 A^{-5/3}$.}
\label{rhfig2}
\end{figure}

\begin{figure}[htb]
 \centerline{\includegraphics[trim=0cm 0cm 0cm
0cm,width=0.45\textwidth,clip]{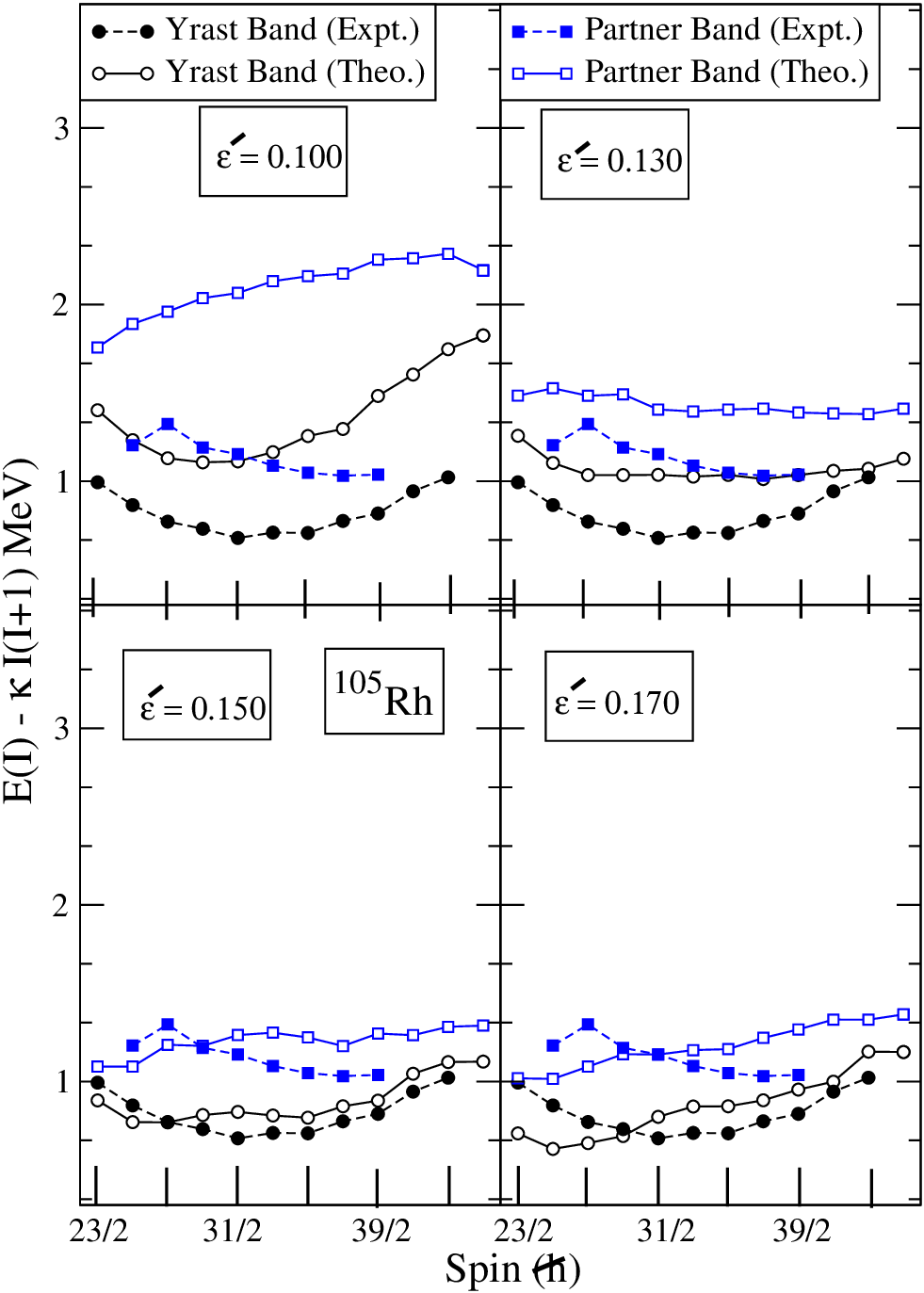}}
\caption{(Color online)  Comparison of the measured energy levels of
of  $^{105}$Rh
 nucleus and the results of TPSM calculations. The
scaling factor $\kappa$ appearing in the y-axis is defined as
$\kappa=32.32 A^{-5/3}$.}
\label{rhfig3}
\end{figure}


\begin{figure}[htb]
 \centerline{\includegraphics[trim=0cm 0cm 0cm
0cm,width=0.45\textwidth,clip]{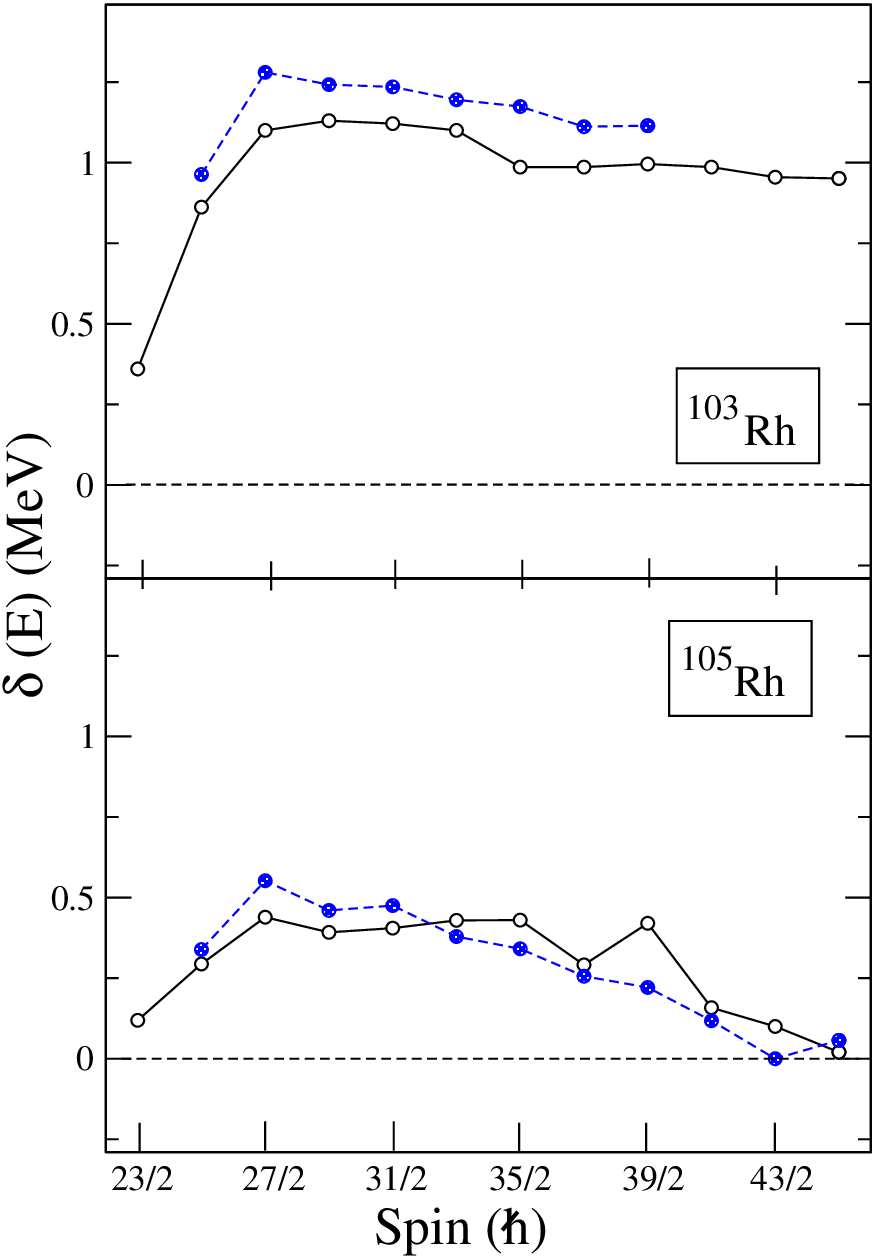}}
\caption{(Color online)  Energy difference, $\delta(E)$, between the 
doublet bands and plotted as a function of spin for the two studied Rh-isotopes.}
\label{rhfig4}
\end{figure}
\begin{figure}[thb]
\begin{center}
\includegraphics[width=0.9\columnwidth]{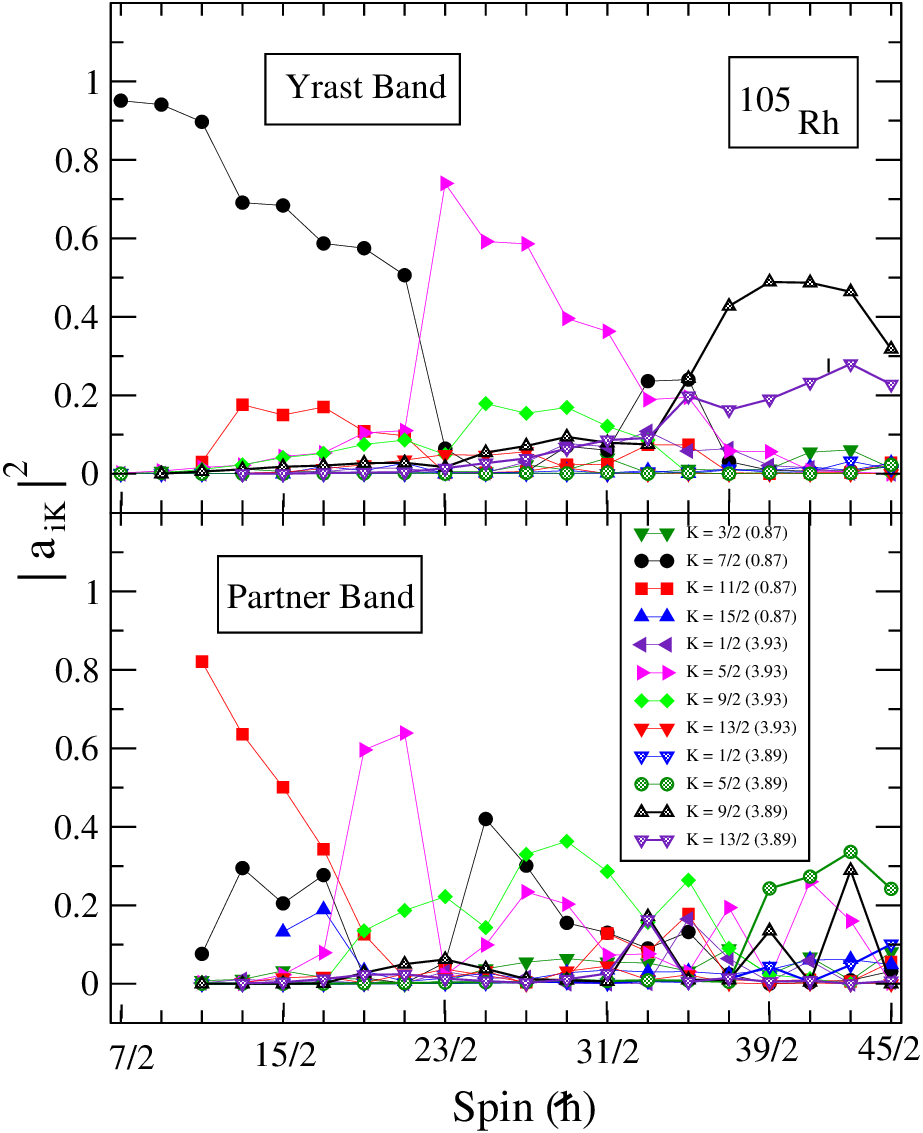}
\end{center}
\caption{ (Color online)
 Probability of various projected K-configurations in the wavefunctions
of the yrast and the first excited bands for $^{105}$Rh.}
\label{rhfig5}
\end{figure}
In recent years, it has been also demonstrated that doublet band structures
observed in some odd-odd and odd-mass nuclei may be a manifestation of
the breaking of the chiral symetry in the intrinsic frame of
reference \cite{VD00,PO04,PRM,RPA1,RPA2}. The chiral symmetry is possible for nuclei having triaxial
shapes with total angular momentum having components along all the three
mutually perpendicular axis. In odd-odd nuclei,
the three angular-momentum vectors that form the chiral geometry are that of
core and of valence neutron and proton. The three angular-momentum
vectors can either form left- or right-handed system and can be
transformed into each other with the chiral symmetry transformation operator,
$\hat{T}~\hat{R}_y$, where $\hat{T}$ is the time-reversal operator and
$\hat{R}_y$ is the
rotation by $180^0$ about y-axis \cite{SK02}.  The chiral dynamical variable
so called ``handedness'' ($\sigma$) is not a measurable quantity and
what is measurable in the laboratory frame is the set of 
states of chiral doublet bands which have a well defined value of
complementary variable, the chirality ($\Sigma$). In the strong chiral
symmetry breaking limit with the three angular-momentun vectors
perpendicular to each other, $\sigma$ assumes the values of $\pm 1$.
The left- and right-handed states are well separated and there is no 
possibility of tunneling between the two states. This results into two 
degenerate doublet bands in the laboratory frame of reference. 

For the weak chiral symmetry breaking, the three  angular-momentum 
vectors are not orthogonal, which as a matter of fact 
is true in most of the physical
situations, and the handedness variable takes the value between +1 and
-1 [the value of 0 corresponds to the planar situation]. For this
case, the tunneling between the two states takes place and in the
laboratory frame this corresponds to the mixing of the two solutions
with the result that two bands tend to be non-degenerate. 

It is evident from the above discussion that a fingerprint of the 
chiral symmetry is the energy difference between the states of the doublet
bands. In Fig.~\ref{chifig1}, this energy difference, $\delta(E)$, is
plotted for odd-odd Cs-isotopes \cite{bh14a} using both observed and
TPSM calculated energies. The doublet band observed in these
nuclei have been proposed to arise from the breaking of the chiral 
symmetry mechanism. It is apparent from the figure that
$\delta(E)$ varies from 0.2 to 0.4 for most of the cases. As already
pointed out above, for the case of strong chiral symmetry breaking
this value should be close to zero and also it should be constant as a
function of angular-momentum \cite{grodner}. It is noted from
Fig.~\ref{chifig1} that $\delta(E)$ varies with angular-momentum, in
particular, for heavier Cs-isotopes and, therefore, it is difficult
to make a definitive statement on the chiral nature of the observed
doublet bands for heavier Cs-isotopes.

Further, it has been demonstrated in Ref.~\cite{grodner} that similar
analysis as that for the energies can be also performed for the
transition probabilities as chiral symmetry operator commutes, 
not only with the Hamiltonian, but also with the electromagnetic
transition operators. The deviation between the yrast and the partner
bands for the transition probabilities B$(\lambda\mu;I_{i}\rightarrow I)$ is defined as \cite{grodner}
\begin{equation}
\epsilon (\lambda\mu;I_{i}) =\frac{A-B}{A+B}  , \nonumber
\label{ctr1}
\end{equation}
where
\begin{eqnarray}
&&A =\sqrt{(2I_{i}+1)B_{yrast}(\lambda\mu;I_{i}\rightarrow I)}  , \nonumber\\
&&B = \sqrt{(2I_{i}+1)B_{side}(\lambda\mu;I_{i}\rightarrow I)}\nonumber.
\label{ctr2}
\end{eqnarray}
The above quantity is displayed in Figs.~\ref{chifig2} and \ref{chifig3}
 for the $E2$
and $M1$ transitions using the TPSM expressions and wavefunctions.
 The figures also
display the experimental values, wherever available. It is evident
from Fig.~\ref{chifig2} that calculated values for $E2$ transitions are close to zero line for most
of the cases and the experimental values available for $^{126}$Cs and
$^{128}$Cs are well reproduced by the TPSM approach. The  deviations
for the $M1$ transitions, depicted in Fig.~\ref{chifig3}, are again noted to
be close to zero line for the light Cs-isotopes. For the
heavier Cs-isotopes, in particular, for $^{128}$Cs, $\epsilon (M1)$ is
somewhat larger and varies with spin.  


For odd-mass nuclei, the three angular-momentum vectors that form
chiral geometry are that of core, odd-proton (neutron)  and a pair of
neutrons (protons). It has been proposed that in odd-proton $^{103}$Rh
and  $^{105}$Rh isotopes that first excited band is the normal
$\gamma$-band up to spin, I=21/2, but above this spin value this band
forms the chiral partner of the yrast band. In order to investigate how
$\gamma$-bands in these isotopes are transformed into the conjectured chiral
bands, we have carried out TPSM study for these isotopes \cite{bh14}
within the space N $=$ (3, 4, 5) for neutrons and  (2, 3, 4) for protons,
and pairing strengths of $G_1=20.25$ and $G_2=16.20$.

As an illustrative example, the band diagram for $^{105}$Rh is
shown in Fig.~\ref{rhfig1} and in order to have a better visualization 
of the bands,
the energies have been subtracted by a core contribution. The
ground-state band has the intrinsic configuration of one quasiparticle
with K=7/2. For odd-mass nuclei, the $\gamma$-band head has two possible
configurations with $K_g=K \pm 2$, resulting into $K_g=11/2$ and
$3/2$. Both these $\gamma$-bands are shown in Fig.~\ref{rhfig1} with
$K_g=11/2$ being favoured in energy. It is noted from the figure that
the ground-state band is crossed at I=19/2 by another band having 
K=5/2 which is
a three-quasiparticle state.  It is also seen that the $\gamma$-band
built on this three-quasiparticle state having K=9/2 also crosses the
ground-state configuration at I=23/2. There is a further simultaneous
crossing by two three-quasiparticle bands having K=9/2 and 13/2 with
intrinsic energy of 3.89 MeV   at I=33/2. 

The projected bands depicted in Fig.~\ref{rhfig1} and many more [around forty
for each angular-momentum] are employed to diagonalize the shell model
Hamiltonian as explained earlier. The bands obtained after band mixing
are shown in Figs.~\ref{rhfig2} and \ref{rhfig3} for $^{103}$Rh and $^{105}$Rh,
respectively. The results of yrast and the first excited band are
only shown above I=23/2$~\hbar$  as they have been proposed to originate
from the chiral symmetry breaking above this spin value. It is evident
from the two figures that only for the optimum triaxial
deformation of $\epsilon'=0.15$ [$\gamma=28^0$ for$^{103}$Rh and $\gamma=33^0$ for$^{105}$Rh], a good agreement
is obtained between theoretical and the experimetal energies. 
In Fig.~\ref{rhfig4}, $\delta(E)$ is displayed as a function of angular
momentum for the isotopes and it is quite evident that
deviation is quite large and, therefore, it is difficult to classify
these doublet bands as chiral partners. 

In order to probe further the evolution of the yrast and the first excited
band structures with spin, the wavefunctions of the two bands
are plotted in Fig.~\ref{rhfig5} for $^{105}$Rh as a representative case.
It is noted from the upper panel that yrast band up to I=21/2$~\hbar$
has the dominant K=7/2 configuration and above this spin value
it is K=5/2 three-quasiparticle configuration that is important. It
is also seen that above I=21/2$~\hbar$, the three quasiparticle configuration
with K=9/2 dominate the yrast states. The first excited band from
I=11/2$~\hbar$ up to I=17/2$~\hbar$ has the expected dominant component
of K=11/2
which is the $\gamma$-band built on the ground-state configuration.
However, above I=17/2$~\hbar$, it is a highly mixed state and ceases to be
a $\gamma$-band.


\begin{figure}[thb]
\begin{center}
\includegraphics[width=0.9\columnwidth]{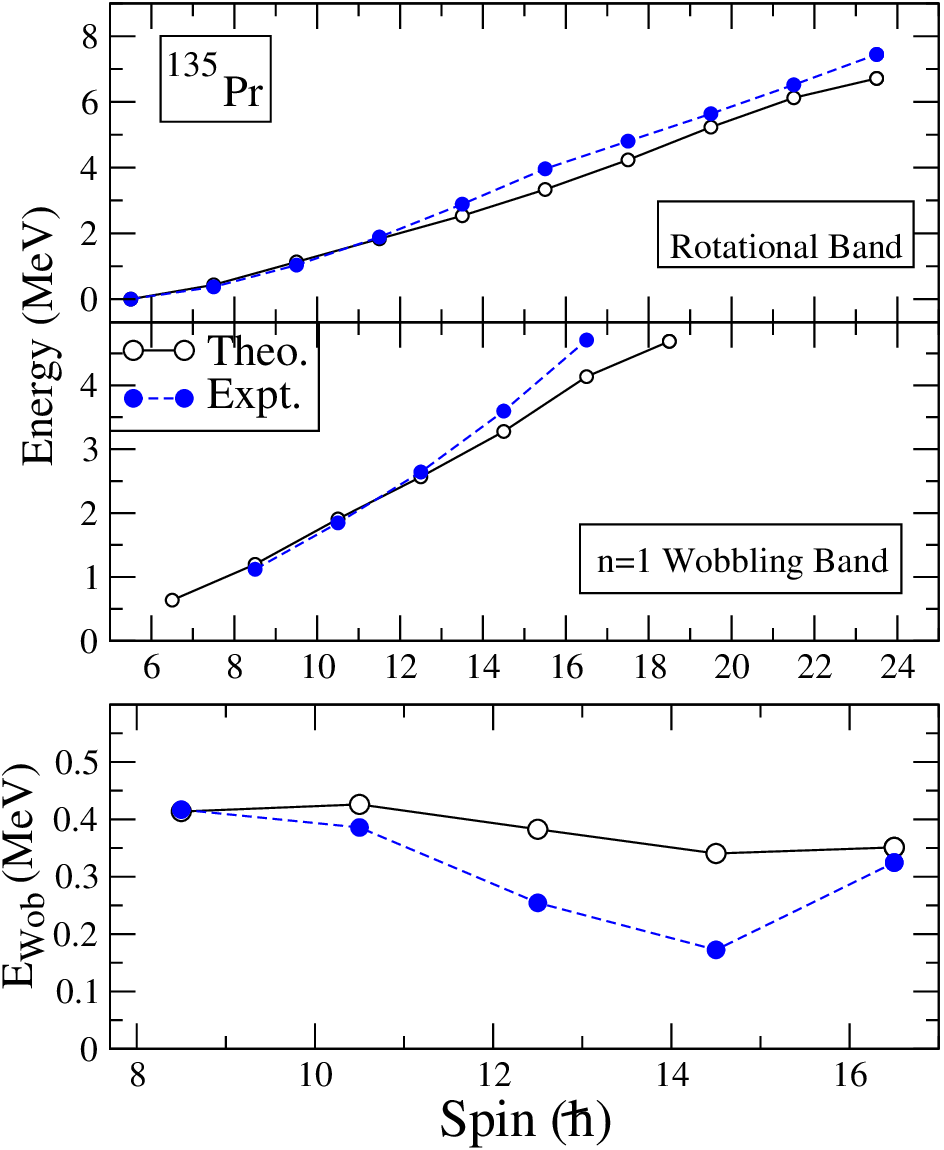}
\end{center}
\caption{ (Color online)
Top two panels depict comparison between the experimental and TPSM
results for the yrast and n=1 wobbling bands in $^{135}$Pr. Bottom
panel shows the comparison of the  variation of wobbling frequency
with spin.
} \label{wbfig1}
\end{figure}
\begin{figure}[thb]
\begin{center}
\includegraphics[width=0.9\columnwidth]{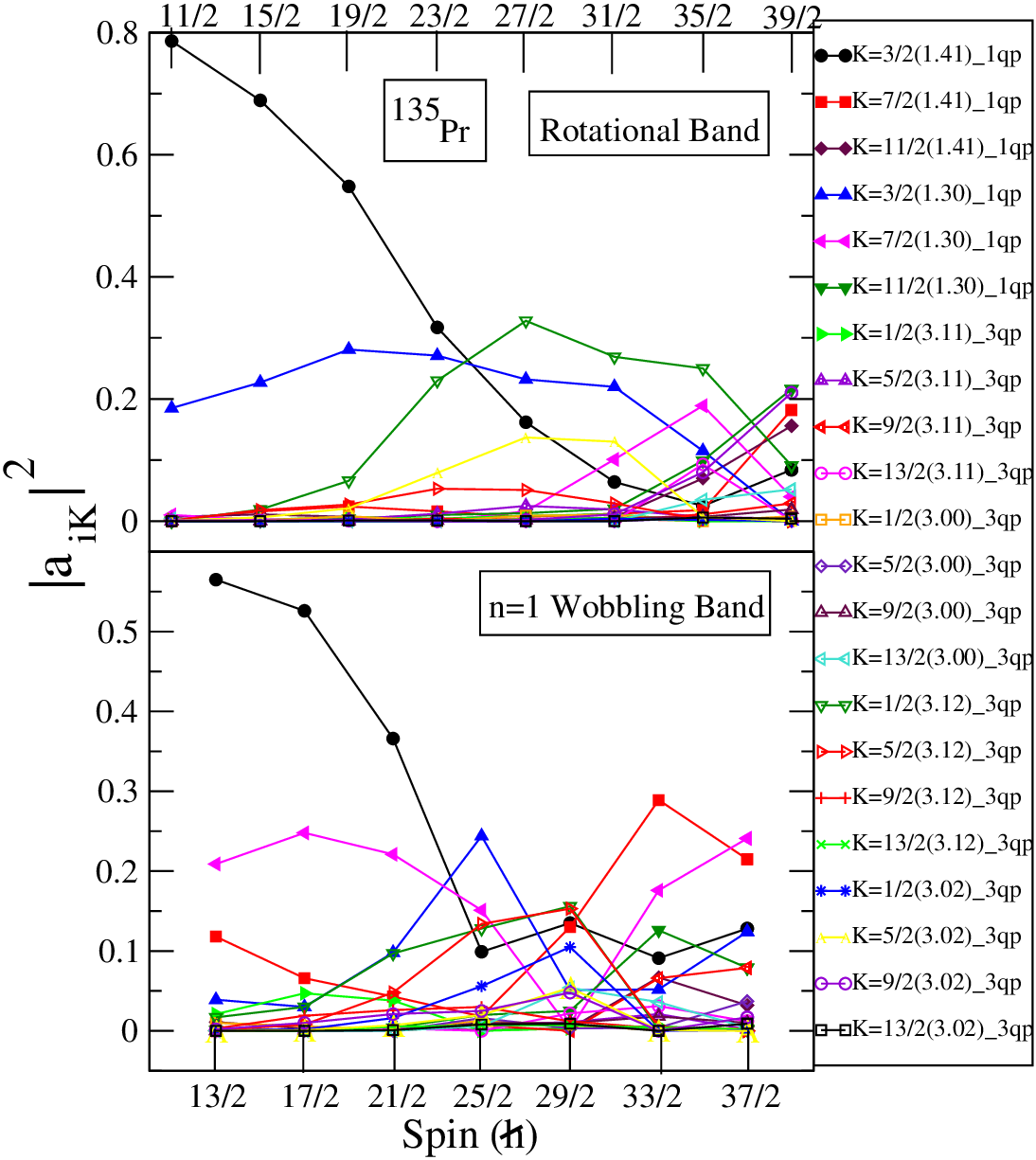}
\end{center}
\caption{ (Color online)
 Probability of various projected K-configurations in the wavefunctions
of the yrast and the wobbling n=1 bands in $^{135}$Pr. The
contributions from various configurations are labelled by $K$,
intrinsic energy and the rank. 
}  \label{wbfig2}
\end{figure}

\subsection{Wobbling Motion}
\label{Sect.04}
Wobbling motion like chiral symmetry, discussed above, is possible only
for triaxial nuclei. This mode was predicted by Bohr and Mottelson in late
sixties \cite{BM75} for even-even nuclei and it was shown that the frequency of the wobbling mode is
proportional to angular momentum, I with enhanced $\Delta~I=1$, E2
inter-connecting transitions. It has been recently shown that for
odd-mass nuclei, the wobbling mode is modified depending on the
direction of the angular-momentum vector of the odd-particle. For
systems with angular-momentum of the odd-particle aligned along the medium axis 
of the core which has largest moment of inertia (referred to as
longitudinal mode) the wobbling frequency decreases with spin. This
frequency increases if angular-momentum of the last
particle is anti-parallel to the medium axis and has been called as the
transverse mode. The wobbling frequency is given by \cite{ma15} 
\begin{equation*}
\begin{split}
 E_{wob}(I)& = E(I, n_\omega =1)\\
  &-\frac{[E(I-1,n_\omega=0)+E(I+1,n_\omega=0)]}{2}.
\end{split}                
\end{equation*}
In the present work we have performed TPSM analysis 
 to describe the
observed band structures and the transverse wobbling mode in $^{135}$Pr 
nucleus. These calculations
have been performed with the deformation parameters, $\epsilon=0.16$
and $\epsilon'=0.11$ within the configuration space of 
N$=3,4,5$ major shells for neutrons and 
N$=2,3,4$ for protons; and with 
the pairing interaction parameters, $G_1=20.12$ and $G_2=13.13$. The results obtained
after diagonalization of the shell model Hamiltonain are presented in
Fig.~\ref{wbfig1}. In the top two panels of the figure, the results are
compared for the yrast and the n=1 wobbling bands. In the bottom
panel, the calculated wobbling frequency calculated from the TPSM energies is
compared with the frequency obtained from the measured energies.
It is evident from the figure that wobbling frequency from
TPSM depicts less drop with spin as compared to the
frequency obtained from the measured energies. 

The wavefunctions for the yrast and the n=1 wobbling bands
are depicted in Fig.~\ref{wbfig2} and it is evident that
three-quasiparticle  band crosses the one-quasipartice at
I=25/2$~\hbar$ for both the bands. It turns out that this crossing occurs much earlier
in $^{133}$La and plays a vital role in understanding the
longitudinal wobbling mode observed recently for this system \cite{SP15}.
The calculations
reported here are preliminary and we are in the process of performing
a detailed 
analysis of the wobbling motion for all the nuclei where it has been
observed. 

\section{Summary and Future Prospects}
\label{Sect.05}
During the last decade, research in nuclear theory has witnessed a
discernable progress in the development of state-of-the-art models
and techniques to elucidate the rich variety of shapes and structures
in nuclei. There is a great optimism that in the
coming years it should be possible to apply these  { \it {Ab-initio} } methods 
to investigate majority of the properties all across the nuclear
periodic table with the availability of more powerful computing
facilities. However, at the moment these methods have limited
applicability and are used to describe nuclei in lighter
mass regions only. To study, for instance, the rich band structures
observed in medium and heavy mass regions, alternative methods with
moderate computational requirements ought to be explored. 

Recently, TPSM approach has been developed to describe the rich band
structures observed in well deformed and transitional nuclei. This
model employs the basis that are solutions of the triaxial Nilsson
potential and then three dimensional projection is performed to
project out the states with well defined angular momentum
quantum number. The
advantage of this approach is that systematic studies of a large class
of nuclei can be performed with a minimal computational effort. As a
matter of fact, already a number of systematic investigations have been
undertaken using this model and it has been demonstrated to reproduce
the known experimental data remarkably well. This model has been applied
to investigate a broad range of properties related to the triaxial
degree of freedom of the nuclear deformation. 

It is known that most of
the deformed nuclei are axially symmetric with well defined angular momentum
projection along the symmetry axis so called the $"K"$ quantum
number. Band structures in well deformed nuclei are labelled with this
quantum number and transitions in these nuclei are known to follow the selection
rules based on this quantum number. However, there are also many regions so
called transitional regions where the axial symmetry is known to be
broken and a triaxial degree of freedom plays an important role. 

In most of these triaxial nuclei, $\gamma$-bands are observed which
traditionally have been considered as vibrations in the non-axial
degree of freedom. In the TPSM interpretation, $\gamma$-bands emerge from the
projection of the K = 2 component of the triaxial vacuum
configuration. 
$\gamma$-bands are known in some Dy, Hf, Hg and U nuclei up to very
high angular momentum and in the present work we have performed a systematic investigation
of these nuclei using the TPSM approach. The observed yrast and $\gamma$-bands in
all these studied nuclei have been demonstrated to be well described
by the TPSM calculations. Deviations have also been noted above I = 22 and a
possible reason for this discrepancy could be due to neglect of the
four-quasineutron and proton configurations in the present work.

The possibility of observing $\gamma$-bands based on the quasiparticle
configurations has also been explored in the present work. It is quite
evident from the very construction of the basis states in TPSM
approach that $\gamma$-bands are possible based on each intrinsic state. 
$\gamma$-band based on the ground-state are quite well established and have
been studied using many different approaches and methods. However, $\gamma$-bands
built on the excited quasipartice configurations have remained rather abstruse
as most of the earlier models didn't consider the quasiparticle
excitations. It has been demonstrated that some of the excited band
structures in Ce- and Nd- isotopes are, as a matter of fact, the
$\gamma$-bands based on two-quasiparticle states. In some of these
isotopes, two s-bands are observed with similar g-factors and this has
remained an unsolved problem for several years. In the conventional
approach, two s-bands are expected to be based on neutron and proton
aligned structures and, therefore, g-factors of the two s-bands should
have opposite signs. Measurements of g-factors of two s-bands provide
same signs for both the s-bands and, thereby, indicating that both the
s-band have either proton or neutron structures. What has been shown in the TPSM studies that second
s-band in many of these nuclei is the $\gamma$-band based on the two-
quasineutron or proton states. As the parent band and the
$\gamma$-band based on it originate from the same intrinsic
configurations, two s-bands in the TPSM approach are expected to have
similar g-factors. TPSM has also povided an explanation  on the
observation of three s-bands in $^{68}$Ge. It
has been shown that one of the s-band is the $\gamma$-band based on
the two-neutron quasiparticle state.

Further, recently new observations have been made in a set of nuclei
that are considered as fingerprints of the triaxial deformation. These
new observations include the occurrence of doublet bands and excited
bands with dominant $\Delta~I=1$ E2 transitions and have been
regarded as manifestaions of chiral and wobbling motion in rotational 
nuclei. These observations are entirely as a consequence of the 
triaxial shape of system and are not  possible in the axial limit. In
the present work, the appearance of the doublet bands have been 
investigated for odd-odd Cs-isotopes from A = 124 to 132 and also 
for the odd-proton $^{103}$Rh and $^{105}$Rh isotopes. It is expected
that the doublet band should be degenerate in the limit of
strong chiral symmetry breaking. It has been shown that the energy difference of the
Cs-isotopes for lighter isotopes is small and, therefore, may be regarded
as originating from the chiral symmetry breaking mechanism. However, for
heavier isotopes, the energy difference is large and also they have
a considerable variation with spin and, therefore, the doublet bands
cannot be considered as chiral partners. We have also evaluated the
differences in the BE2 and M1 transitions and similar conclusions
have been  drawn from these quantities as from the energy considerations.

The high-spin doublet band structures in $^{103}$Rh and $^{105}$Rh have also
been proposed to originate from chiral symmetry breaking. In the
low-spin regime, these nuclei have regular $\gamma$-bands and in the
high-spin region, the observed energy difference between the $\gamma$-band and
the yrast sequence decrease with spin and based on this inference, the
high spin band structures have been regarded as chiral partners. [For
the normal $\gamma$-band, this difference
remains almost constant]. It has been shown that energy difference,
both experimental and theoretical, is too large as compared to the
Cs-isotopes and for some angular momentum value it is more than $1~$MeV.
It has been also demonstrated in our earlier study \cite{JY01} that
  quadrupole moment varies with spin quite appreciably and, therefore,
the doublet bands in $^{103}$Rh and $^{105}$Rh cannot be regarded
as candidates for chiral symmetry breaking. 

We would like to add that one of the the major problem in the TPSM
model is that first of all the pairing plus quadrupole interaction
used is quite rudimentary and needs to be generalised for a better
desciption of the nuclear properties.  Secondly, the coupling
constants of the interaction are
adjusted through self-consistency condition with the input deformation
parameters.  This means that optimum deformation value for a system
under investigation should be known prior to performing the TPSM
calculations. Although, the final results should be independent of the
input deformation in case a very large basis set is employed in the study,
however, in practice a limited basis space is employed and the results
become basis dependent. In order to circumvent this problem, we are
considering to fix the coupling constant of the interaction through a
mapping procedure as has been done in other approaches  \cite{Alh08,Alh06}.
The energy surface, for instance, obtained from the density functional
theory can be mapped to the surface obtained using the model
interaction used in the TPSM approach with adjustable coupling
constants.

There are several other possible ways of improving the predictive
power of the TPSM calculations. For instance, the diagonalization of the 
shell model Hamiltonian is performed for a fixed value of the
deformation parameter and for a more accuate description, a set of
deformation values should be considered by employing the generator
coordinate method (GCM). This will allow
the possibility to  have coupling between the $\gamma$-deformation and
$\gamma-$vibrational degrees of freedom. One of the major discrepancy
that has surfaced in the application of the TPSM approach is the band
head energy of the $\gamma\gamma$-band \cite{GH08,JG09,SB10}.
 It has been observed
that for most of known
cases, TPSM calculations underpredicts band head energy of this band by more
than 1 MeV. It is expected that $\gamma\gamma$-band to have
siginificant vibrational component because of mixing with the
quasiparticle states which are close in energy. We hope that by
performing GCM with both $\beta$ and $\gamma$ as generator coordinates, the
$\gamma\gamma$-band and other properties shall be described more accurately.
It is also quite important to include higher
multi-quasiparticle states in the basis space. It has been noted in the 
present investigation that TPSM results tend to disagree above I = 22
and the reason for this discrepancy is the neglect of the four-proton and 
-neutron quasiparticle states. We are presently working to improve
the TPSM approach along the lines discussed above and the results shall
be presented in future publications. 

\section*{Appendix}
The matrix elements of the Hamiltonian and transition operators in
TPSM approach are evaluated using the Wick theorem. For instance, for 
a two-body operator of the form $\hat O^{\dagger}\hat O$, we have
\begin{eqnarray*}
 \hat O^{\dagger} \hat O&&={\langle \hat O\rangle}^2+\langle \hat O\rangle \{ :\hat 
O^{\dagger}:+:\hat O:\}+:\hat 
O^{\dagger} \hat O:\\
&& \equiv \hat H^{(0)} +\hat H^{(1)}+\hat H^{(2)} 
\end{eqnarray*}
where $\langle \hat O\rangle$ is the contraction and $:\hat O:$ is the
normal ordered form of the operator. The rotated matrix 
elements of one-body and two-body operators are given by
\begin{eqnarray*}
&&\langle \hat H^{(1)}[\Omega]\rangle = \langle \hat O\rangle \{ \langle :\hat O^{\dagger}:[\Omega]\rangle+\langle 
:\hat 
O:[\Omega]\rangle \}\\
&&\langle \hat H^{(2)} [\Omega]\rangle = \langle :\hat 
O^{\dagger}:[\Omega]\rangle\langle :\hat 
O:[\Omega]\rangle
\end{eqnarray*}
where $[\Omega]= \hat R(\Omega) / \bra\hat R(\Omega)\ket $.
The detailed expressions of the above rotated matrix elements are given in Ref.~\cite{KY95}.
In TPSM approach, diagonalization of the shell model Hamiltonian
follows from the Hill-Wheeler method. In this method, the following
ansatz is used for wavefunction
\begin{equation}
  |\psi\ket = \int d\Omega ~F(\Omega)~\hat R(\Omega)~|\phi \ket \label{a1b4}
 \end{equation}
For the projection operator, the expansion coefficient in the above
equation is written as
\begin{equation*}
 F(\Omega)=\sum_{IMK}\frac{2I+1}{8\pi^2}F^I_{MK}D^I_{MK}(\Omega)
\end{equation*}
Using the variational ansatz
\begin{equation*}
 \delta\big[\frac{\bra\psi|\hat H|\psi\ket}{\bra\psi|\psi\ket}\big]=0,
\end{equation*}
and substituting $|\psi\ket$ from Eq.~(\ref{a1b4}), we obtain
\begin{equation}
 \sum_{\kappa^{'}K^{'}}\{\mathcal{H}_{\kappa K \kappa^{'}K^{'}}^{I}-E\mathcal{N}_{\kappa K 
\kappa^{'}K^{'}}^{I}\}F^{I}_{\kappa^{'} K^{'}}=0, \label{a15}
\end{equation}
 where the Hamiltonian and norm kernels are given by
 \begin{eqnarray*}
 && \mathcal{H}_{\kappa K \kappa^{'}K^{'}}^{I} = \langle \phi_{\kappa}|\hat H\hat 
P^{I}_{KK^{'}}|\phi_{\kappa^{'}}\rangle ,\\
&&\mathcal{N}_{\kappa K \kappa^{'}K^{'}}^{I}= \langle \phi_{\kappa}|\hat P^{I}_{KK^{'}}|\phi_{\kappa^{'}}\rangle .
 \end{eqnarray*}
In the TPSM model, we  work in a representation in which the norm matrix is diagonal, i.e.,
 \begin{equation*}
  \sum_{\kappa^{'}K^{'}}\mathcal{N}_{\kappa K 
\kappa^{'}K^{'}}^{I}U^{\sigma}_{\kappa^{'}K^{'}}=n_{\sigma}U^{\sigma}_{\kappa K},
 \end{equation*}
and solve Eq.~(\ref{a15}) with the eigenstates of the above norm equation
as the basis states.

\section*{Acknowledgement}
We would like to express our deep gratitude to S. Frauendorf, Y. Sun,
R. Palit and R.N. Ali for their contributions at various stages in the
development of the triaxial projected shell model approach.


\end{document}